\documentclass[fleqn,usenatbib]{mnras}

\usepackage[T1]{fontenc}
\usepackage{graphicx}	
\usepackage{amsmath}	
\usepackage{amssymb}	
\usepackage{txfonts}
\usepackage{bm}
\usepackage{comment}
\usepackage{gensymb} 
\usepackage{orcidlink}
\usepackage{etoolbox}

\title[Planet--disc interactions around binaries]{
Planet-disc interactions around eccentric binaries and misaligned ring formation}

\author[R. G. Martin \& S. H. Lubow]{Rebecca G. Martin$^{1,2}$\orcidlink{0000-0003-2401-7168} and Stephen
  H. Lubow$^3$\orcidlink{0000-0002-4636-7348}\\ $^{1}$Nevada Center for Astrophysics, University
  of Nevada, Las Vegas, 4505 South Maryland Parkway, Las Vegas, NV
  89154, USA \\ $^{2}$Department of Physics and Astronomy, University
  of Nevada, Las Vegas, 4505 South Maryland Parkway, Las Vegas, NV
  89154, USA \\ $^3$Space Telescope Science Institute, 3700 San Martin
  Drive, Baltimore, MD 21218, USA}

\date{}
\pubyear{2025}

\begin{document}
\label{firstpage}
\pagerange{\pageref{firstpage}--\pageref{lastpage}}
\maketitle

\begin{abstract}
  We explore the evolution of a  giant planet  that interacts with a circumbinary disc   that orbits a misaligned binary by means of analytic models and hydrodynamical simulations.  Planet-disc interactions lead to mutual tilt oscillations between the planet and the disc. Even if circumbinary gas discs form with an isotropic mutual misalignment to the binary, planet-disc interactions can cause giant planets to evolve towards coplanar or polar alignment.     For a low-mass disc, the binary dominates the dynamical evolution   of the planet  leading to a wide range of circumbinary planet inclinations. For a high-mass disc, the disc dominates the dynamical evolution of the planet and planet inclinations move towards coplanar or polar alignment to the binary orbit, depending upon the initial disc inclination  and the binary eccentricity.     In addition, for a high-mass disc ($\sim 50\, M_{\rm p}$) and a high initial disc inclination, the planet can undergo Kozai--Lidov oscillations that can result in the planet being ejected from the system. For initially highly misaligned systems, the non-coplanarity of the planet and the disc can lead to long-lived inner misaligned disc rings that can become highly eccentric.
\end{abstract}


\section{Introduction}

Observations of circumbinary discs show that they can be misaligned to
the binary orbital plane
\citep[e.g.][]{Winn2004,Chiang2004,Capelo2012,Kennedy2012,Brinch2016,Aly2018,Kennedy2019}. The
misalignment may be a result of chaotic accretion during the star
formation process
\citep[e.g.][]{McKee2007,Bate2003,Monin2007,Bateetal2010,Offner2010,Tokuda2014,Bate2018, Elsender2023} or
stellar flybys \citep[e.g.][]{Clarke2001,Cuello2019b,Nealon2019flyby,Smallwood2023}. Discs around
binaries with short orbital periods, less than about $30\,{\rm days}$,
have aligned discs while those with longer orbital periods have a
larger range of inclinations and binary eccentricities
\citep{Czekala2019}.

A misaligned disc around a circular orbit binary undergoes uniform
nodal precession as the angular momentum vector of the disc
precesses about the angular momentum vector of the binary. If the disc
is sufficiently narrow and warm it may precess as a solid body
\citep[e.g.][]{Larwoodetal1996,LP1997,Young2023,Rabago2024}. The viscosity in the disc leads to dissipation
and thus the disc aligns towards the binary orbital plane
\citep{PT1995,PL1995,Lubow2000,Nixonetal2011b,Nixon2012,Kingetal2013,Facchinietal2013,Lodato2013,Foucart2013,Foucart2014}.

Studies of ballistic circumbinary particles suggest that a misaligned circumbinary disc around an eccentric orbit binary with
sufficiently high tilt can undergo nodal libration oscillations of the
tilt angle and longitude of ascending node \citep{Verrier2009,
  Farago2010,Doolin2011, Naoz2017, Chen2019}.  A low mass disc around an eccentric orbit binary can reside in a polar configuration in which the disc
 is perpendicular to the binary orbital plane with the angular momentum vector of the disc aligned to the binary eccentricity vector \citep{Aly2015, Martin2017, Zanazzi2018,Cuello2019,Rabago2023}. In this state, the disc does not precess in the inertial frame.  For
higher mass discs around eccentric orbit binaries, there is a generalised polar state  in which the disc does not precess relative to the binary. This state occurs at lower tilt
angles \citep{MartinLubow2019,Chen2019,Chen2020,Abod2022}.

When the disc is librating, the angular momentum vector  precesses about the generalised polar alignment \citep{Martin2019}. A warm protostellar circumbinary disc undergoing libration can evolve towards polar alignment due to the effects of disc dissipation \citep{Martin2017, Martin2018, Lubow2018, Zanazzi2018,Franchini2019b,Cuello2019,Smallwood2020}.  A circumbinary disc around an eccentric
binary with low initial misalignment also undergoes tilt oscillatons
and non-uniform precession as it aligns to the binary orbital plane
\citep{Smallwood2019}. The fraction of polar aligned circumbinary discs may be significant around isolated binaries \citep{Ceppi2024,Johnson2025}, although companion stars to the binary reduce the parameter space over which polar alignment occurs \citep{Martin2022,Ceppi2023,Lepp2023,Lepp2025}. 

Giant planets form while the gas disc is present \citep[e.g.][]{Lagrange2010}
and thus we expect their initial alignments to be similar to the
disc. The Kepler mission has observed more than 10 circumbinary planets, all of
which are close to coplanar  with the binary
\citep{Doyle2011,Welsh2012,Orosz2012a,Orosz2012b,Kostov2014,Welsh2015,Li2016,Kostov2016}.  The binaries are roughly uniformly distributed in mass ratio \citep{MartinDV2019b}. All except one of the Kepler detected circumbinary planets orbit around a low eccentricity binaries. The exception is Kepler-34b that has a binary eccentricity of 0.52 \citep{Welsh2012,Kley2015}. More recently, radial velocity methods have been used to find circumbinary planets \citep{Triaud2022,Standing2023,Baycroft2024}.   Long-lived highly inclined protoplanetary circumbinary discs can orbit around eccentric orbit binaries, but are unlikely to occur around circular orbit binaries. The low binary eccentricities in the Kepler sample are in turn due to their short binary orbital periods that cause these binaries to be subject to substantial stellar  tidal dissipation of their eccentricities
\citep[see ][]{Raghavan2010}. Consequently,
the observed near coplanarity in the Kepler cases is  likely due to selection effects, since these binaries generally have
relatively short periods and low binary eccentricity \citep[e.g.][]{MartinDV2015,MartinDV2017,MartinDV2019}. Recently, evidence of retrograde apsidal precession of a binary has been observed \citep{Baycroft2025}, the most likely explanation is that it is driven by a polar circumbinary planet \citep[e.g.][]{Zhang2019, Childs2023,Lepp2023,Lubow2024}. Polar planets may form in a polar circumbinary disc as efficiently as coplanar planets in a coplanar disc \citep[e.g.][]{Childs2021,Childs2021b}.

All of the observed circumbinary planets are near the limit of dynamical stability close to the binary  \citep{Holman1999,Sutherland2016,Quarles2016,Quarles2018,Hong2019}. Planet formation close to the binary may be difficult due to excitation of planetesimal velocities \citep{Paardekooper2012,Meschiari2012,Meschiari2014,Lines2014,Silsbee2015}.  The  planetesimal eccentricity growth may be suppressed by a gas disc \citep{Rafikov2013,Martin2013} or else planetary migration may be required to form the observed planets  \citep[e.g.][]{Nelson2003,Pierens2007,Pierens2008,Pierens2013,Kley2014,Mutter2017,Kley2019,Yamanaka2019}.  The Kepler data
may contain transiting planets around non-eclipsing binaries and thus
in the future more inclined circumbinary planets may be found
\citep{MartinTriaud2014,MartinDV2017}. Circumbinary planets may also be found with binary eclipse timing variations \citep[e.g.][]{Getley2017,Zhang2019,Socia2020},  
transit timing variations and transit duration variations \citep{Windemuth2019}, microlensing \citep{Bennett2016,Luhn2016}, and astrometry \citep{Sahlmann2015}.

The planet formation process in a circumbinary disc is altered by the
torque from the binary that is not present in the single star case
\citep[e.g.][]{Nelson2000,Mayer2005,Boss2006,Martinetal2014b,Fu2015,Fu2015b,Fu2017,Franchini2019}. 
Consider a giant planet that is initially embedded in a gaseous disc in a binary star system and is sufficiently massive to open a gap in the disc.
The planet and disc are initially mutually coplanar but inclined relative to the binary.\footnote{See \cite{Childs2023} and \cite{Lubow2024} for examples of  planet dynamics in the case of initially misaligned planet orbits.}
In the case of a giant planet embedded in a circumstellar disc, previous studies have shown that such a configuration may not remain coplanar
\citep{Picogna2015,Lubow2016,Martin2016,Pierens2018,Franchini2020}. 
The planet and disc undergo mutual circulation if the disc mass
is sufficiently low, resulting in mutual tilt oscillations with amplitude larger than the initial tilt relative to the binary. 
In this paper we are concerned with the evolution of such an initial configuration involving a circumbinary planet and disc.
A planet that is misaligned to a disc may cause warping and dynamical evolution of the disc \citep[e.g.][]{Matsakos2017,Arzmasskiy2018,Nealon2018,Zhu2019}.  

In Section~\ref{analytic} we determine the secular evolution of a
mildly inclined circumbinary planet--disc system around a circular or an eccentric
orbit binary. In Section~\ref{sims} we describe the 3D smoothed particle
hydrodynamic (SPH) simulations that include the viscous evolution of
the disc that interacts with the planet and the binary star system. We describe the results of these simulations for circular orbit binary systems in 
Section~\ref{circular}  and for eccentric orbit binaries
in Section~\ref{eccentric}.
In Section~\ref{concs} we draw our conclusions and  discuss the implications of our results.

\section{Secular evolution of nearly coplanar circumbinary planet--disc systems}
\label{analytic}

We apply analytic methods \citep{Lubow2000, Lubow2016} to
describe the secular evolution of a circumbinary planet--disc system around a binary of eccentricity, $0 \le e_{\rm b} \le 1$.
We assume that planet and disc  tilts are small relative to the binary orbital plane.
We apply a secular torque based on \cite{Farago2010} that is determined in the quadrupole
approximation away from the binary. We assume the disc warping is small so that in lowest
order we can take the disc to be flat. This approximation is valid provided that the sound
crossing time across the disc is short compared to the nodal precession timescale \citep{Larwoodetal1996,LP1997}. The disc 
and planet are assumed to reside on nearly circular orbits. We neglect effects of disc self-gravity
and the effects of the planet and disc  on the orbit of the binary.  We also neglect any
density evolution of the disc. The binary orbit is fixed, we neglect the effects of general relativity \citep[but see][]{Lepp2022,Childs2024bh} and tides \citep[e.g.][]{Chen2024}. 

\subsection{Secular Model}
\label{secular}

We apply a Cartesian coordinate system $(x,y,z)$ in which the origin is the binary center of mass and
reference plane is the $x-y$ plane.
The binary lies in the $x-y$ plane with its angular
momentum vector along the positive $z$ direction and
the binary eccentricity vector $\bm{e}$ lies along the positive $x-$direction.
Tilt evolution is described by tilt vectors that are unit vectors $\bm{\ell}_{\rm p}(t)$ and $\bm{\ell}_{\rm d}(t)$
that are along the direction of the planet and disc angular momentum, respectively. 
Under the  approximation that the planet and disc lie close to the $x-y$ reference plane,  it follows that for the planet $|\ell_{{\rm p}x }| \ll 1$, $|\ell_{{\rm p} y}| \ll 1$, $|\ell_{{\rm p}z }|\simeq1$
and similarly for the disc.
Since the tilt vectors are unit vectors, the evolution of only two components needs
to be analyzed, the $x$ and $y$ components. Since the equations that we apply are linear in the tilt vectors, we can take the time dependence of these vectors to
be of the form $\exp{( i \omega t )}$.
 The tilt vector components are complex quantities whose actual values are taken
to be their real parts, in the usual way. For example,  $\ell_{{\rm p}x} (t) =  Re[ \ell_{{\rm p}x } \exp{( i \omega t )} ]$. 
With this approach, there are four eigenmodes, each with a value
of $\ell_{{\rm p}x}, \ell_{{\rm p}y}, \ell_{{\rm d}x}, \ell_{{\rm d}y},$ and $\omega$. Since the equations are linear, we are free to select a normalization that we take as $\ell_{{\rm p}x} =1$. In addition, the tilt vectors need to satisfy initial conditions at $t=0$.

The gravitational interaction between planet and disc 
is described  by a coupling coefficient denoted by $C_{\rm pd}$.
The coupling coefficient  given by
\begin{equation}
C_{\rm pd}=2\pi \int_{R_{\rm in}}^{R_{\rm out}} G M_{\rm p} R \, \Sigma(R) \, K(R,a_{\rm p})\,dR,
\label{Cpd}
\end{equation}
where the symmetric kernel, with units of inverse length, is given by
\begin{equation}
K(R_j,R_k)=\frac{R_j R_k}{4\pi} \int_{0}^{2\pi} \frac{ \cos \phi \,d\phi}{(R_j^2+R_k^2-2R_j R_k\cos \phi)^{3/2}}.
\label{K}
\end{equation}
The Keplerian orbital
frequency around the binary of mass $M$ is $ \Omega_{\rm K}(R) = \sqrt{G\, M/R^3}$. 
The angular momentum of the planet   that orbits at radius $R_{\rm p}$  is (approximately) given by $J_{\rm p}  =  M_{\rm p} R_{\rm p}^2 \Omega_{\rm K} ( R_{\rm p})$.
For a disc that extends radially from some inner radius $R_{\rm i}$ to an outer radius $R_{\rm o}$, its angular momentum  is given by
$J_{\rm d}  = 2 \pi \int^{R_{\rm o}}_{R_{\rm i}} \Sigma(R) R^3 \Omega_{\rm K}(R) dR$.

We are interested in both circular orbit and eccentric orbit binaries.
The nodal precession frequency about a circular orbit binary  plays a useful role
in calculations of the eccentric binary case as well.
The nodal precession frequency of a particle in a nearly coplanar orbit about a circular orbit binary is given by
\begin{equation}
\omega_{\rm c}(R)= \frac{3}{4} \frac{M_1 M_2}{M^2} \Omega_{\rm b} \left(\frac{a}{R} \right)^{7/2}
\label{omc}
\end{equation}
\citep[e.g.][]{Farago2010},
where the binary semimajor axis is denoted by $a$, the binary component masses are $M_1$ and $M_2$ with $M=M_1+M_2$
and the binary orbital frequency is $\Omega_{\rm b}$.
Therefore, the nodal precessional frequency of a slightly inclined planet about a circular orbit binary
is given by 
\begin{equation}
\omega_{\rm p} =  \omega_{\rm c}(R_{\rm p}).  \label{omegap}
\end{equation}
The nodal precessional frequency of a slightly inclined disc about a circular orbit binary
is given by the angular momentum weighted average of precession frequencies  
\begin{equation}
\omega_{\rm d} = \frac{2 \pi \int_{Ri}^{Ro}   \omega_{\rm c}(R) \Sigma(R) R^3 \Omega_{\rm K}(R) \, dR}{J_{\rm d}} \label{omegad}
\end{equation}
\citep{PT1995}.

The evolution equations for the planet tilt $\bm{\ell}_{\rm p}(t)$ and the disc tilt $\bm{\ell}_{\rm d}(t)$ are given by 
\begin{eqnarray}
i \omega J_{\rm p}  \ell_{{\rm p}x } &=&  C_{\rm pd} ( \ell_{{\rm p}y} -  \ell_{{\rm d}y}) + \tau_x  \omega_{\rm p} J_{\rm p}  \ell_{{\rm p}y},  \label{lpx}\\
i \omega J_{\rm p}  \ell_{{\rm p}y } &=&  C_{\rm pd} ( \ell_{{\rm d}x} -  \ell_{{\rm p}x}) + \tau_y   \omega_{\rm p} J_{\rm p}  \ell_{{\rm p}x},  \label{lpy}\\
i \omega J_{\rm d}  \ell_{{\rm d}x } &=&  C_{\rm pd} ( \ell_{{\rm d}y} -  \ell_{{\rm p}y}) + \tau_x \omega_{\rm d} J_{\rm d}  \ell_{{\rm d}y},  \label{ldx}\\
i \omega J_{\rm d}  \ell_{{\rm d}y } &=&   C_{\rm pd} ( \ell_{{\rm p}x} -  \ell_{{\rm d}x}) + \tau_y  \omega_{\rm d} J_{\rm d}  \ell_{{\rm d}x}.  \label{ldy}
\end{eqnarray}
Quantities  $\tau_x$ and $\tau_y$ are related to the secular torque on the planet and
disc due to a circular or eccentric orbit binary with eccentricity $e_{\rm b}$. We apply the torque given by equations 2.1.6 - 2.1.8 in \cite{Farago2010}. 
In the case that  planet and disc are nearly coplanar with respect to the binary,
we have that
\begin{eqnarray}
\tau_x   &=&  \pm(1 - e_{\rm b}^2),  \label{txco} \\
\tau_y  &=&  \mp(1 + 4 e_{\rm b}^2),  \label{tyco} 
\end{eqnarray}
where  the upper (lower) sign on the right side is taken when the object (planet or disc) is in prograde (retrograde) 
orbit about the binary. We consider only the prograde case in this paper.

These equations are solved analytically in Mathematica.
First, with normalization condition $\ell_{{\rm p}x} =1$, 
Equations (\ref{lpx}) - (\ref{ldy}) are solved for $\ell_{{\rm p}y}$, $\ell_{{\rm d}x}$, $\ell_{{\rm d}y}$, and $\omega$
in terms of  $C_{\rm pd}$,  $J_{\rm p}$,  $J_{\rm d}$, and $e_{\rm b}$. The solution consists of four eigenmodes.
Next, the initial conditions are imposed to determine the weights of the modes that are used to determine the unique solution. Each mode is represented
as a column in a 4x4 matrix ${\cal M}$ in the form of the modal solutions for $(\ell_{{\rm p}x}, \ell_{{\rm p}y}, \ell_{{\rm d}x}, \ell_{{\rm d}y})^T$, where $T$
denotes the transpose operator that transforms the row vector into a column vector.
The weights $w_j$ for mode number $j=1, 2, 3, 4$  are represented in column vector $w$ that is computed by
\begin{equation}
w = {\cal M}^{-1} \ell_0
\label{w}
\end{equation}
where $\ell_0$ is the column vector  that contains the known initial  tilts $(\ell_{{\rm p}x0}, \ell_{{\rm p}y0}, \ell_{{\rm d}x0}, \ell_{{\rm d}y0})^T$.
Weight column vector $w$ is determined analytically through the matrix inversion.
The full solution for the tilts in time is then given by
\begin{equation}
\ell(t) = Re[ {\cal M} \, w(t)],
\end{equation}
where $\ell(t)$ is the solution column vector  $(\ell_{{\rm p}x}(t), \ell_{{\rm p}y}(t), \ell_{{\rm d}x}(t), \ell_{{\rm d}y}(t))^T$ 
and  $w(t)$ is the column vector $(w_1 \exp{(i \omega_1 t)},..., w_4 \exp{(i \omega_4 t)})^T$ with the subscripts denoting the mode number.

In evaluating results, we calculate the  planet and disc inclination with respect to the reference plane as
\begin{equation}
i_{\rm j}(t) = \sqrt{ \ell_{{\rm j}x}^2(t) +  \ell_{{\rm j}y}^2(t)} \,  i_{\rm j0},
\label{i}
\end{equation}
where $j= \,$p, d and $i_{\rm j0}$ is the initial tilt value.
We determine the longitude of ascending node of the planet and disc as
\begin{equation}
\phi_{\rm j}(t) = \arctan{(- \ell_{{\rm j}y}(t),  \ell_{{\rm j}x}(t))},
\label{phi}
\end{equation}
where again $j=\,$p,d.
We determine the relative inclination between  the planet and disc that begin at a common inclination $i_0$ as
\begin{equation}
\Delta i (t)  = \sqrt{ (\ell_{{\rm p}x}(t)-\ell_{{\rm d}x}(t))^2+  (\ell_{{\rm p}y}(t)-\ell_{{\rm d}y}(t))^2} \, i_0.
\label{Deltai}
\end{equation}

 We consider an equal mass binary.
 The  planet mass is taken to be $M_{\rm p}=0.001 \,M$, which opens a gap in the disc.
 The planet is located at $R_{\rm p} = 5a$, where $a$ is the binary semi-major axis and is assumed to lie
 interior to the circumbinary disc.  
 The disc inner edge initially extends from $R_{\rm i}=7 a$ and follows a power law as $\Sigma(R) \propto R^{-3/2}$. This clearance between the planet and disc is typical of what
 is found in simulations involving a $1 M_{\rm J}$ planet \cite[e.g.,][]{Lubow2016}.
 We typically take the disc outer edge to be $R_{\rm o}= 10 a$, but consider other values as well.
  We typically take the disc mass to be $M_{\rm d} = 0.01 M$, but consider other values as well.
 We take the initial condition to be that the disc and planet are mutually aligned but tilted about the $y$-axis,
 corresponding to a longitude of ascending node $\phi = -90^{\circ}$. 
 The initial conditions used in Equation (\ref{w}) correspond to planet and disc tilt vectors that are along the negative $x-$direction
 \begin{equation}
 \ell_0 = (-1,0,-1,0)^T.
\end{equation}
Since the equations are linear in the tilts, we can choose to normalize the initial tilts in this way.
But, the tilts should in reality be small enough to satisfy the approximation we are making. This normalization is valid, provided all results on tilts
are expressed relative to the initial tilt.

\begin{figure}
\begin{centering}
\includegraphics[width=0.8\columnwidth]{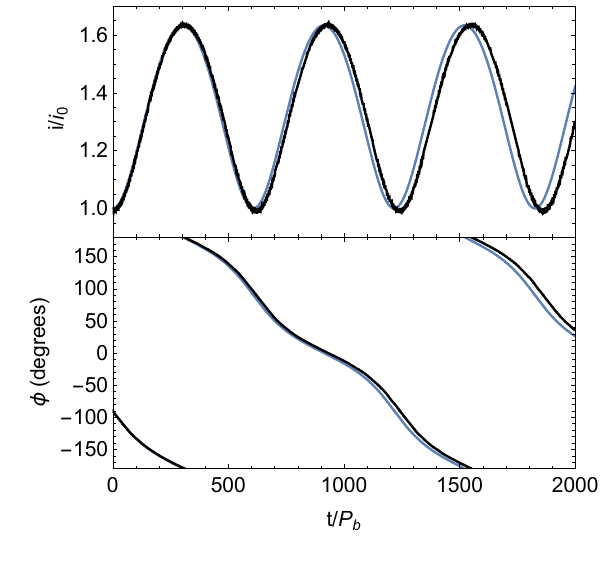}
\end{centering}
\caption{Comparison between the evolution predicted by the analytic secular model (blue) and numerical orbit integration (black) for a planet that initially orbits
at radius $R=5 a$ from the centre of mass of an equal mass binary that has eccentricity $e_{\rm b}=0.5$. The upper panel plots evolution of the orbit inclination from the binary orbital plane
relative to the initial value. The lower panel plots the evolution of longitude of the ascending note. Time is given in units of the binary orbital period. }
\label{particletest}
\end{figure}

As a check on the accuracy of the secular model, we consider the case of a planet-disc system with parameters described above, but with a negligible
disc mass. We take the binary eccentricity to be $e_{ \rm b}=0.5$. We compare the secular evolution of the planet 
to that obtained by numerically integrating the orbit of a test particle that has the same starting conditions as the planet
in the secular model. The particle begins on a circular orbit that is tilted by $i_0=10^{\circ}$ with respect to the binary orbital plane.
The tilt and longitude of ascending node of the particle as a function of time are determined by its instantaneous tilt vector.
As seen in Fig.~\ref{particletest}, the agreement between the two methods of calculation is quite good. The changing tilt and nonuniform precession
are due to the binary eccentricity. There is a small but growing difference between the results due to a $2\%$ difference in the precession periods.

\subsection{Mutual planet-disc circulation and libration}

\begin{figure}
\begin{centering}
\includegraphics[width=\columnwidth]{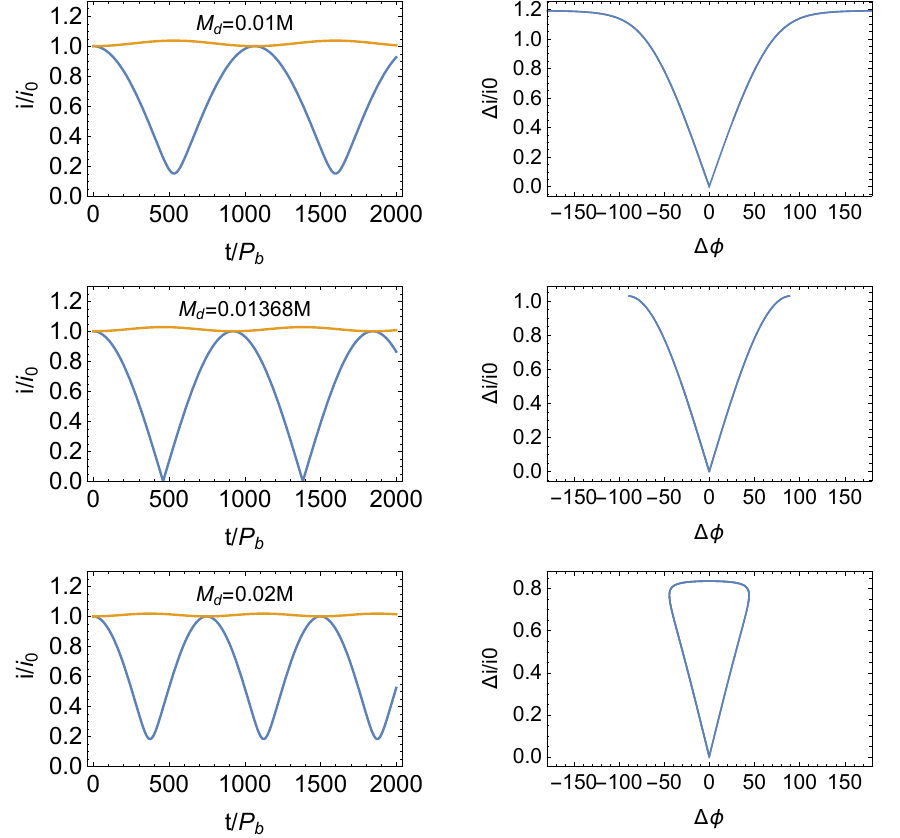}
\end{centering}
\caption{Inclination evolution of a planet and disc that are nearly coplanar with a central circular orbit binary ($e_{\rm b}=0$) for parameters given at the beginning of Section~\ref{secular}, but for different values of disc mass in units of the binary mass.
Left column: evolution of the tilt relative to the binary orbital plane for the planet (blue) and disc (orange).
Right column: phase portrait of the relative tilt  between the planet and disc as a function of their nodal phase difference measured in degrees.
The lowest (highest) mass disc case undergoes circulation (libration). The middle case is at the critical mass between circulation and libration 
that occurs because the minimum tilt of the planet is zero.}
\label{coplanarcirc}
\end{figure}

\begin{figure}
\includegraphics[width=0.8\columnwidth]{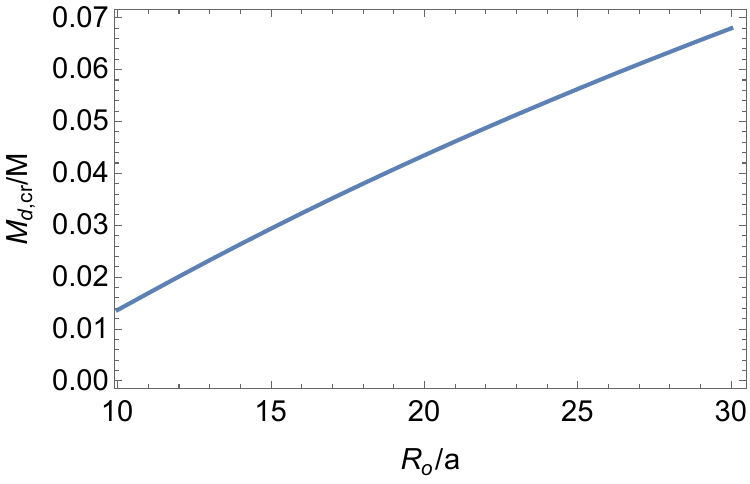}
\caption{ Critical disc mass $M_{\rm d,cr}$ (in units of the binary mass) between libration and circulation for a circular orbit binary in a system with parameters given in Section~\ref{secular},
but with different values of the disc outer radius, $R_{\rm o}$, in units of the binary semi-major axis.  }
\label{coplanarcircMcr}
\end{figure}

We consider the conditions under which the planet and disc undergo mutual circulation versus libration in  Fig.~\ref{coplanarcirc}. The left panels show the inclination of the disc and the planet relative to the binary as a function of time. The right panels plot phase portraits of the relative planet-disc tilt given by Equation (\ref{Deltai}) as a function of the nodal phase difference using Equation (\ref{phi}). For a low mass disc,  the planet and the disc undergo mutual circulation (top panels). The planet and disc are decoupled to the extent that their nodal
phase difference monotonically increases in time. For a high mass disc (bottom panels), the planet and the disc undergo mutual libration. The planet and disc are coupled
together by their mutual gravitational interaction to the extent that the nodal phase difference
is limited  and oscillates about zero.
In the case of a circular orbit binary, the tilt
of the planet relative to the binary orbital plane is less than or equal to the tilt of the disc for the adopted parameters.
The disc undergoes mild tilt oscillations, while the planet undergoes much larger tilt oscillations. For a small mass disc (top panels) the minimum planet tilt decreases with increasing disc
mass while the planet and disc undergo mutual circulation.

At a critical disc mass, $M_{\rm d,cr}$, the minimum planet tilt is zero, as shown in the middle panels.
At this critical disc mass, the system is between mutual libration and circulation.
For this disc mass, both components of the planet tilt change sign at the time that $\ell_{\rm p}$ goes through zero, resulting in a $180^\circ$ phase shift.
As seen in the middle right panel, the relative phase is limited to $\pm 90^\circ$, unlike the circulating case
in the top right panel. This phase range of $180^\circ$ is due to the planet phase shift when $\ell_{\rm p}$ goes through zero.
The phase portrait
does not form a closed loop, as in the librating case of the bottom right panel.
The overall pattern is similar to our previous results for a planet-disc system that
orbits a member of a circular orbit binary star system \citep{Lubow2016}.
This critical mass is an important indicator of the type of interaction that occurs
between the planet and disc.

The condition for the transition between mutual circulation and libration occurs when
the system properties are such that for  $i_0>0$,
\begin{equation}
    \min_t [ i_{\rm p}(t)]= 0.
    \label{librcrit} 
\end{equation} We note that a different and equally valid condition was given in \cite{Lubow2016},
but condition (\ref{librcrit}) is easier to calculate.
The condition is a consequence of the fact that the planet nodal phase given by equation (\ref{phi})  undergoes a change in behaviour for systems in which both of the planet tilt components go through zero simultaneously.

Fig.~\ref{coplanarcircMcr} plots the critical disc mass between mutual circulation and mutual libration of the planet and the disc, $M_{\rm d,cr}$, as a function of disc
outer radius, $R_{\rm o}$ with other parameters given at the start of this section. We see that as the
outer radius increases, the critical disc mass increases. The reason is that
the strength of disc interaction with the planet is dominated by the mass of the disc near its inner edge.
Keeping that disc mass roughly constant while increasing its outer radius requires an increasing disc mass.

\begin{figure}
\begin{centering}
\includegraphics[width=\columnwidth]{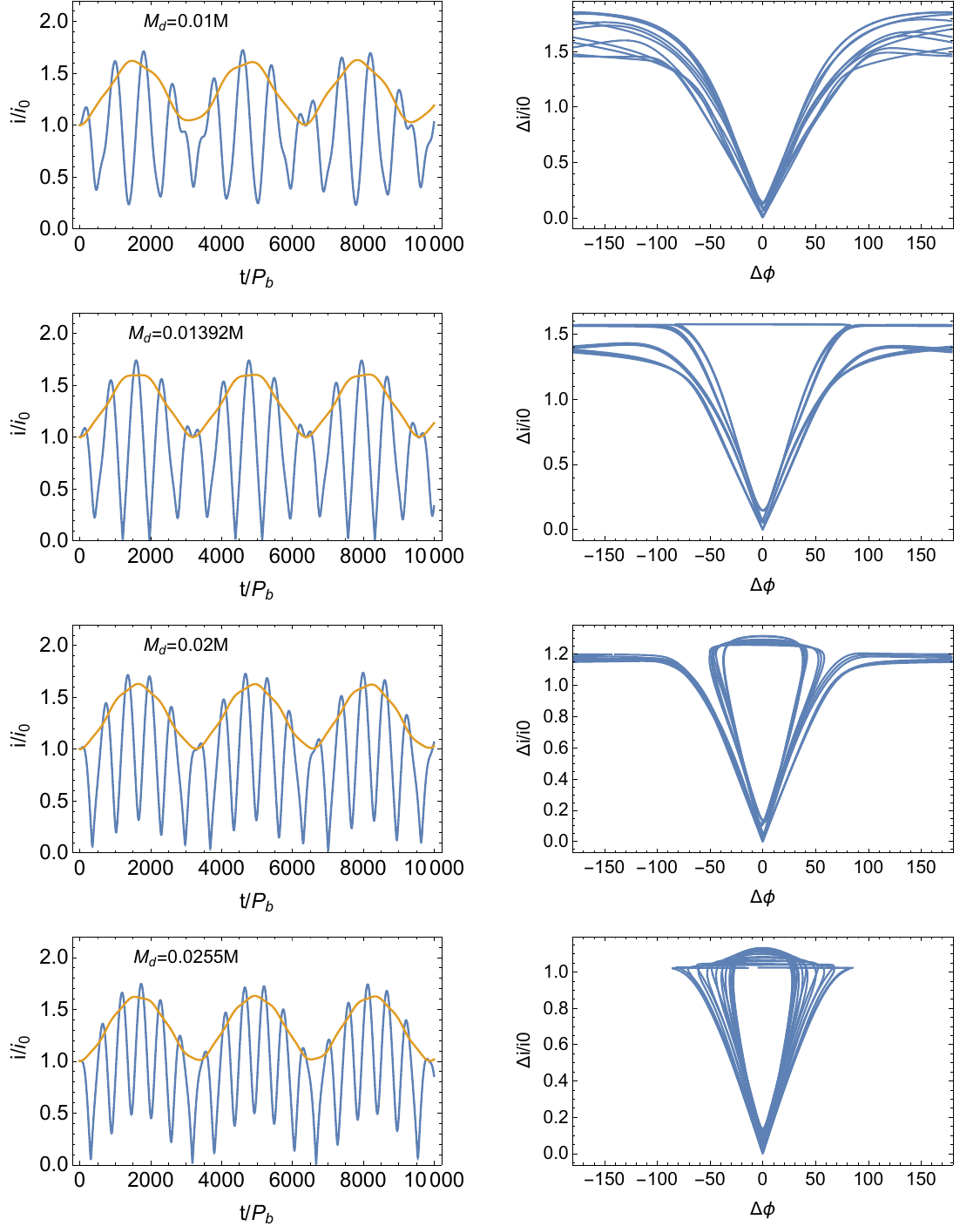}
\end{centering}
\caption{ Inclination evolution of a planet and disc that are nearly coplanar with central eccentric orbit binary for  ($e_{\rm b}=0.5$)  for parameters given at the beginning of Section \ref{analytic},
but for different values of disc mass in units of binary mass that increase from top to bottom.
Left column: evolution of binary tilt relative to the binary orbital plane for the planet (blue) and disc (orange).
Right column: phase portrait of the relative tilt  between the planet and disc as a function of their nodal phase difference.
The top row undergoes undergoes only circulation. The second row from the top is at the critical 
disc mass at which the system begins to undergo libration over the plotted time interval. The third case undergoes both circulation and libration. The bottom row is at the disc mass that begins to undergo only libration.}
\label{coplanarecc5}
\end{figure}

Fig.~\ref{coplanarecc5} is similar to Fig.~\ref{coplanarcirc}, but involves a  binary
with eccentricity $e_{\rm b}=0.5$. In the eccentric binary orbit case, the evolution is more complicated.
The results in  Fig.~\ref{coplanarecc5} show that the disc tilt oscillations have a much larger amplitude
than in the circular orbit case. In addition, the planet tilt is not always smaller 
than the disc tilt. The planet undergoes oscillations of changing amplitude
over time. 
In the eccentric orbit binary case, there is not a sharp transition between circulation and libration as occurs in the circular binary orbit case.
The lowest mass disc case (top panel) undergoes circulation, but with changing relative tilt $\Delta i$ in different cycles (right panel).
The next disc mass is the smallest disc mass in which the disc
begins to partially librate over the plotted time interval. 
At the next higher disc mass of $0.02 M$, the system exhibits both circulation and libration.
While at the highest disc mass in the bottom panel, the system
has the minimum disc mass such that only librations occur
over the plotted time range. The disc masses over which
the system transitions from circulation to libration in this eccentric binary case ($M_{\rm d}=0.01399  - 0.025 M$) is larger than the critical disc mass in the circular orbit  binary case
($M_{\rm d} = 0.01368 M$).

\begin{figure}
\begin{centering}
\includegraphics[width=\columnwidth]{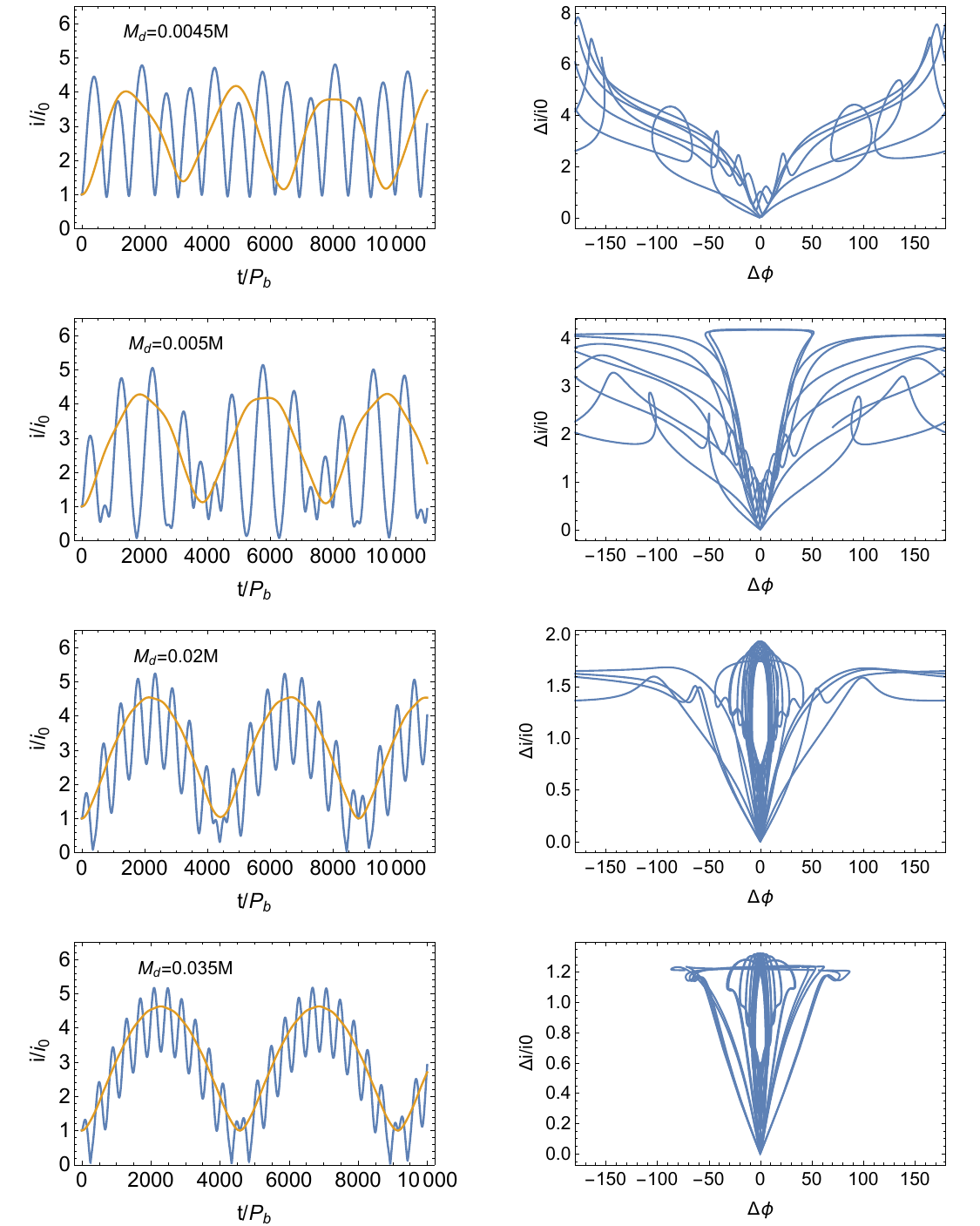}
\end{centering}
\caption{ Similar to Fig.~\ref{coplanarecc5} but with a central eccentric orbit binary for  $e_{\rm b}=0.9$. }
\label{coplanarecc9}
\end{figure}

\begin{figure}
\begin{centering}
\includegraphics[width=0.8\columnwidth]{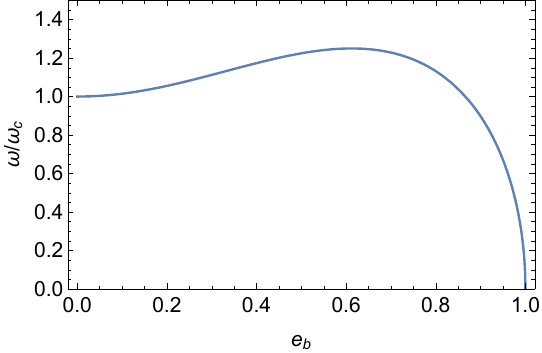}
\end{centering}
\caption{Nodal precession rate of a circumbinary
test particle that is nearly coplanar with  an eccentric orbit binary
binary as a function of binary eccentricity. The precession rate
is normalised by the nodal precesion rate about a circular orbit binary.
}
\label{omprec}
\end{figure}

At a higher binary eccentricity $e_{\rm b}=0.9$, a similar
pattern emerges, but with different disc masses (Fig.~\ref{coplanarecc9}).  In this case, the disc masses over which
the system transitions from circulation to libration in this eccentric binary case 
begins for $M_{\rm d}= 0.005 M$ that is lower than the corresponding value for either the circular binary
orbit case or the less eccentric binary orbit case $e_{\rm b}=0.5$.
This result may be a consequence of the fact that the nodal precession
rate of a test particle  in orbit about a binary varies
the geometric mean of $\tau_x$ and $\tau_y$ in Equations (\ref{txco})
and (\ref{tyco}). This frequency is given by
\begin{equation}
    \omega(r) = \sqrt{1+3 e_{\rm b}^2 - 4 e_{\rm b}^4} \, \omega_{\rm c}(R),
\end{equation}
where $\omega_{\rm c}$ is given by equation (\ref{omc}).
Precession rate $\omega(R)$ is a nonmotontic function of $e_{\rm b}$.
This precession rate is maximum for 
$e_{\rm b} = \sqrt{3/8} \approx 0.612$. As seen in Fig.~\ref{omprec}, $\omega$ for $e_{\rm b}=0.9$
is less than $\omega$ for $e_{\rm b}=0$ which is less than 
 $\omega$ for $e_{\rm b}=0.5$. This ordering is consistent
 with the critical masses we find that are required for mutual (planet-disc)  libration to first set in. For example, the disc mass required for mutual libration in
the $e_{\rm b}=0.9$ case occurs is smaller than the other cases because the disc mass required
for the disc gravity to cause the planet to precess together with it
is smaller. 
This occurs because the planet and disc precession rates are smaller than the other cases we have considered.
On the other hand, the disc mass required to suppress circulations
is higher for the $e_{\rm b}=0.9$ case than the other cases.
This may be a consequence of the larger tilt amplitudes for this case.

\begin{figure}
\begin{centering}
\includegraphics[width=\columnwidth]{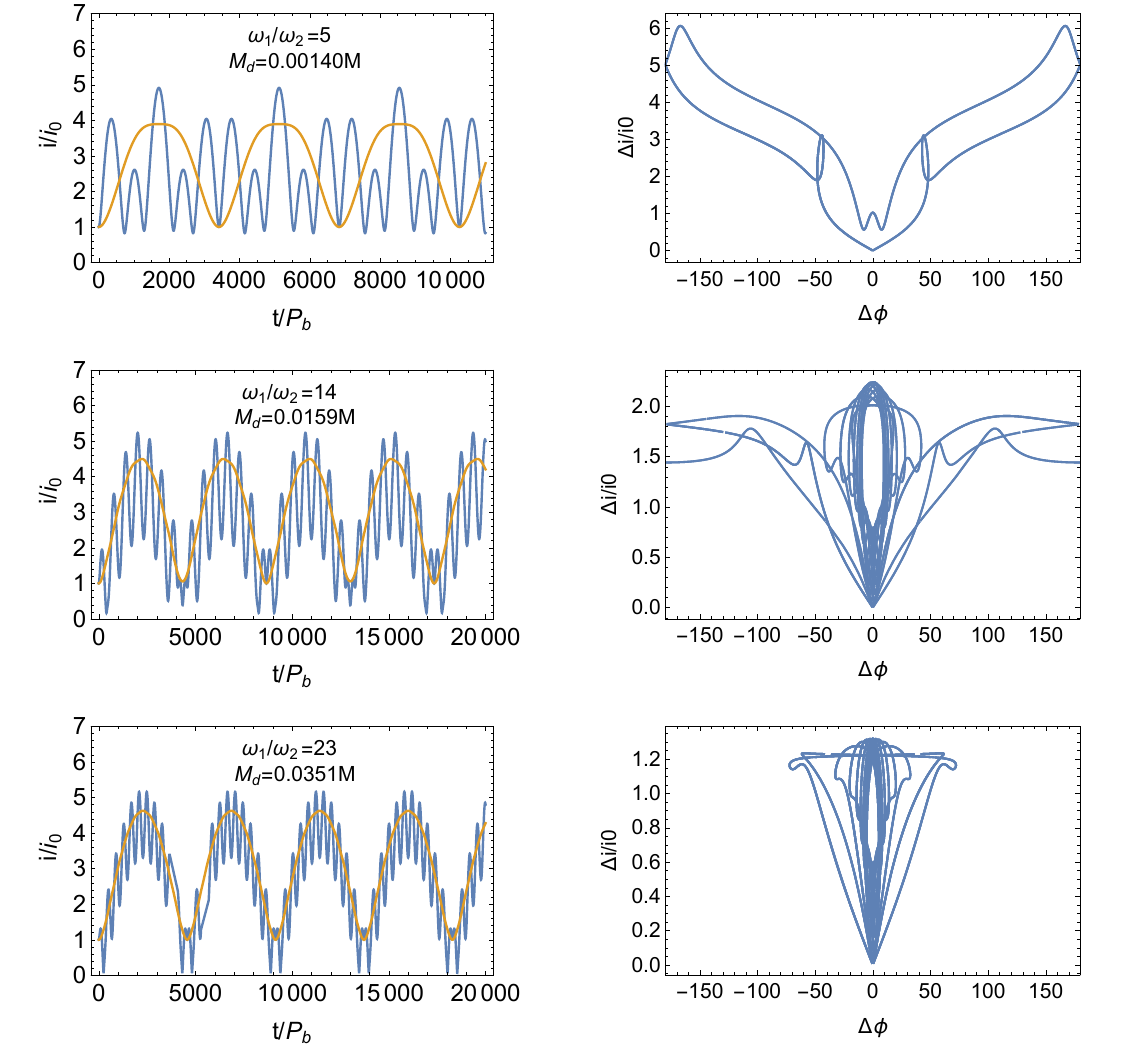}
\end{centering}
\caption{ Similar to Fig.~\ref{coplanarecc9}, also with $e_{\rm b}=0.9$, but these cases all have periodic inclinations with integer mode frequency ratios
given as $\omega_1/\omega_2$ for disc masses given as $M_{\rm d}$ normalised by the binary mass.}
\label{inclperiodic}
\end{figure}

The tilt evolution solution involves a linear combination of the four eigenmodes and eigenfrequencies
for Equations (\ref{lpx})-(\ref{ldy}). Since the model is dissipationless and therefore time-reversible,
for each eigenfrequency $\omega_{i}$ , its negative $-\omega_{i}$ is also an eigenfrequency.
Thus, there are only two different  eigenfrequencies in absolute value.
Usually the ratios of these frequencies are irrational.
As a result, the inclination oscillations are not periodic, except in certain special cases.

In the circular orbit binary, the circular symmetry leads to inclinations variations $i$ at only one frequency (Fig.~\ref{coplanarcirc})
 and is therefore always periodic.
In the absence of the disc, the planet inclination $i_{\rm p}(t)$ oscillates
at a single frequency if the binary is eccentric.
Similarly, the disc inclination $i_{\rm d}(t)$ also oscillates at a single frequency in the absence of the planet. This inclination frequency is simply twice their 
respective nodal precession frequency.
Due to their mutual interactions, the planet and disc
undergo additional oscillations in inclination $i_{\rm p}$ and $i_{\rm d}$  that involve more frequencies.
Consequently, the inclination evolution is generally not periodic, although
an approximate periodicity occurs due to the dominance of the inclination contribution at
twice the nodal precession frequency.

Periodicity of relative tilt as a function of the phase difference 
(phase portraits in right column of Fig.~\ref{coplanarecc5})
also does not generally occur for eccentric orbit binaries, unlike the case of a circular orbit binary. 
Even if the planet and disc do not interact,
the phase portrait is generally not periodic provided that the
binary orbit is eccentric, while the planet and disc tilts $i$ are each periodic in time in that case.

Periodicity of inclination can occur for special rational values of
the modal frequency ratios. Fig.~\ref{inclperiodic} shows
some cases for a binary with $e_{\rm b}=0.9$ in which these ratios are integers. The plot shows cases
for mutual circulation (top panels), a combination of mutual circulation and libration
(middle panels), and for mutual libration (bottom panels). The phase portraits take a simpler form than in the general irrational
ratio cases of Fig.~\ref{coplanarecc9}).

Therefore, in the eccentric orbit binary case, 
the inclination oscillations are generally nonperiodic, although they can be approximately periodic.
Additional nonperiodic behaviour will also occur due to the disc 
density evolution and binary evolution that is not included in this analytic model.
The transition from circulation to libration is not sharp and occurs over a range of disc mass. A significantly higher disc mass is required for the system to be purely librating than in the circular orbit binary case. This occurs because the binary 
eccentricity induces additional tilt oscillations on the planet and disc.

\section{Hydrodynamical simulations} 
\label{sims}

\begin{figure}
\begin{centering}
\begin{center}
\includegraphics[width=0.4\textwidth]{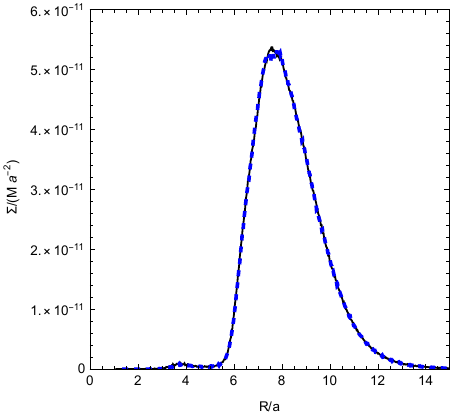}
    \includegraphics[width=0.45\textwidth]{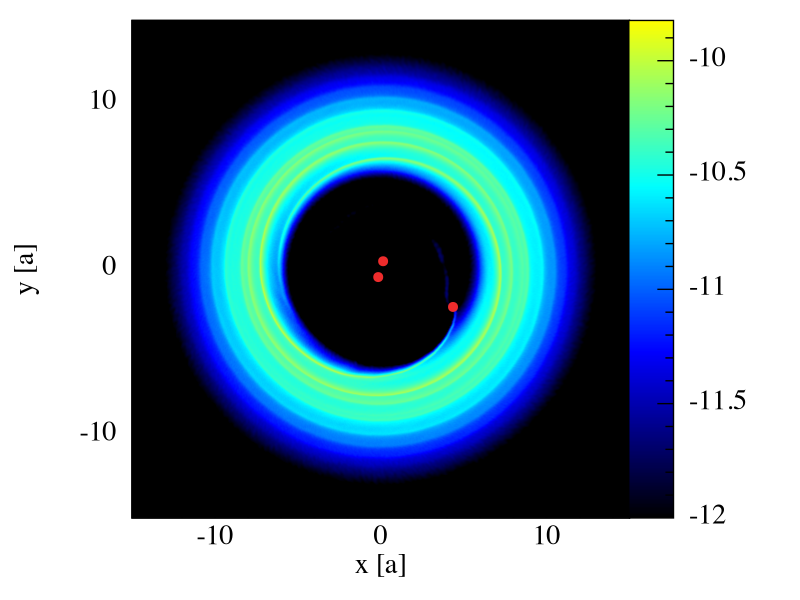}
\end{center}
\end{centering}
\caption{Upper panel: The initial surface density profile for the simulations for
  the circular orbit binary (thin black solid line, coplanar1) and the
  eccentric binary (thick blue dashed line, coplanar2). Lower panel: The disc column density (on a log scale in units of $M \, a^{-2}$) in the orbital plane of the binary (coplanar1).  The red circles show the binary stars and the planet with their radius a factor of 3 times larger than the sink size.  }
\label{sd}
\end{figure}
 
\begin{table*}
\centering
\caption{Parameters of the initial circumbinary disc set up for a binary
  with total mass $M$ and semi-major axis $a$. The planet is initially of mass $M_{\rm p}=0.001\,M$ in a circular orbit around the binary with separation $a_{\rm p}=5\,a$. The first column shows the name of the simulation. The second column shows the figure that the simulation appears in. The third column shows the initial disc mass. The fourth column shows the initial disc inclination with respect to the binary orbital plane. The fifth column shows the initial binary eccentricity. The sixth column describes the relative precession between the planet and the disc. The disc and the planet may
  be in a circulating (C) or librating (L) state with respect to each
  other. The seventh and eighth columns describe whether the planet and disc, respectively, precess about the binary angular momentum vector (C) or the polar state (L).  } 
\begin{tabular}{lcccccccc}
\hline
Name & Figure       & $M_{\rm d}/M$ & $i/^\circ$& $e_{\rm b}$ &
Planet--disc & Planet--binary & Disc--binary\\
\hline
\hline
coplanar1 & \ref{sd} & $1\times 10^{-8}$ & 0 & 0.0 & - & - &  -  \\
coplanar2 & \ref{sd}  & $1\times 10^{-8}$ & 0 & 0.5 & - & - &  -  \\
\hline
circ1 & \ref{sph1}  &  0.001 & 10 & 0.0 & C & C & C  \\
circ2 & \ref{sph1}  &  0.01 & 10 & 0.0 & C & C & C \\
circ3 & \ref{sph1}   &  0.05 & 10 & 0.0 & L  & C & C \\
\hline
circ4  & \ref{sph3}  &  0.001 & 40 & 0.0 & C & C & C  \\
circ5 &  \ref{sph3} &  0.01 & 40 & 0.0 & C & C & C \\
circ6 & \ref{sph3}           & 0.05 & 40 & 0.0 & L & C & C  \\
\hline
ecc1 &  \ref{sph2} &  0.001 & 10 & 0.5 & C & C & C \\
ecc2 &  \ref{sph2} &  0.01 & 10 & 0.5 &  C & C & C \\
ecc3 &  \ref{sph2}  & 0.05 &10 & 0.5 & L & C & C \\
\hline
ecc4 & \ref{sph7}  &  0.001 & 40 & 0.5 & C & L & C  \\
ecc5 &  \ref{sph7}  &  0.01 & 40 & 0.5 & C & C & C \\
ecc6 &  \ref{sph7}  & 0.05 &40 & 0.5 & C & C & C \\
\hline
ecc7 & \ref{sph9}& 0.001 & 60 & 0.5 & C & L & L  \\
ecc8 & \ref{sph9} & 0.01 & 60 & 0.5 & C & L & L \\
ecc9 & \ref{sph9} & 0.05 & 60 & 0.5 & - & - & - \\
\hline
ecc10 & \ref{sph10}  & 0.001 & 88.5 & 0.5 & C & L & L \\
ecc11 &  \ref{sph10}  & 0.01  & 81.2 & 0.5 & C & L & L \\
\hline
\end{tabular}
\label{tab}
\end{table*} 

In this section, we now use 3D hydrodynamical
simulations to consider how the viscous evolution of the disc affects the disc-planet system around a binary. We use
the smoothed particle hydrodynamics code {\sc Phantom}
\citep{LP2010,PF2010,Price2012a,Nixonetal2013,Price2018}.  We model an equal mass binary with mass $M$ and semi-major axis $a$. During the evolution, the binary mass changes by $<0.1\,\%$ and the separation by $<1\,\%$. Changes to the binary eccentricity are generally small, unless otherwise stated.  We follow
the methods of \cite{Lubow2016} and \cite{Martin2016} to find the
initial surface density distribution of the disc with an equilibrium
gap around the planet. We begin with a very low mass disc with mass
$M_{\rm d}=10^{-8}\, M$ and a planet with initial orbital radius
$5\,a$.   The disc mass is
sufficiently low as to not affect the orbital properties or masses of
the binary or the planet. The surface density of the disc is initially
distributed with $\Sigma \propto R^{-3/2}$ between $R_{\rm in}=6\, a$
and $R_{\rm out}=10\, a$. We do not include material interior to the
planet orbit since it is accreted on a short timescale.  

The disc and
the planet are coplanar to the binary orbit. The accretion radius
around each star and planet are $0.1\,a$. Any particles that move inside this radius are considered to be accreted on to the body and their mass and angular momentum are added to the body. Since this accretion radius is much larger than the radius of a planet, this can lead to artificially high accretion rates on to the planet, particularly for high disc mass simulations. 
 The
disc is locally isothermal with aspect ratio $H/R=0.02$ at $R=R_{\rm
  in}$ and sound speed $c_{\rm s}\propto R^{-3/4}$. The \cite{SS1973}
$\alpha$ parameter along with the smoothing length $\left<h\right>/H$ are
constant over the radial extent of the disc \citep{LP2007}. We take
the viscosity parameter $\alpha=0.01$. The disc viscosity is
implemented by adapting the SPH artificial viscosity according to the
procedure described in \cite{LP2010} using $\alpha_{\rm AV} = 0.32$
and $\beta_{\rm AV} = 2.0$. Initially there are $2\times 10^6$
particles in the disc. The disc is resolved with shell-averaged
smoothing length per scale height $\left<h\right>/H=0.31$. 

We run the simulation
until the planet has cleared out its orbit, for 200 orbits of the
planet \citep[e.g.][]{Bateetal2003}. We consider two binary eccentricities, $e_{\rm b}=0$ and $e_{\rm b}=0.5$. These are shown  as coplanar1 and coplanar2, respectively, in Table~\ref{tab}. The resulting surface density
profiles are shown in  Fig.~\ref{sd}. The solid black line in is the circular orbit binary (coplanar1) and the dashed blue line is the eccentric binary (coplanar2). At this
time there are about $1.8\times 10^6$ particles left in the disc.  We
take this distribution of particles and increase the mass to the
desired disc mass and tilt the planet-disc system to the
required inclination. Finding the initial surface density profile in this way limits the effects of the large accretion radius of the planet since the planet gap is cleared before accretion on to the planet can take place. 

In analysing the simulations we divide the disc up into 100 radial bins in spherical radius. In each bin we find the mean properties such as surface density, inclination, longitude of ascending node. Where we show disc evolution we take the bin at radius of $10\,a$.  This corresponds to the outer parts of the disc, outside of the planet orbit. This part of the disc remains quite flat with inclination variables of only a few degrees over the radial extent.  In some cases material is able to flow in past the planet and form an inner ring that is misaligned to the outer ring. In these cases, the disc evolution that we plot corresponds to the outer ring only. 
In Section~\ref{circular} we consider circular orbit binaries and in Section~\ref{eccentric} we consider eccentric orbit binaries.

\section{Circular orbit binaries}
\label{circular}

We first consider a total of 6 simulations around  circular orbit binaries with
different values for disc mass and initial inclination (see circ1--circ6 in
Table~\ref{tab}). For a circular orbit binary, there are no librating
solutions for a third body with respect to the binary. Thus, in all
cases, both the planet and the disc must circulate with respect to the
binary (see the last two columns in Table~\ref{tab}). However, the planet and the disc may be
circulating or librating with respect to each other (column~6). Circulating means
that they nodally precess on independent timescales. Librating means that they nodally
precess on the same average timescale.

\subsection{Low-inclination disc}
\label{circlowinc}

\begin{figure*}
\begin{centering}
\textbf{Analytic model}\par\medskip
\includegraphics[width=1.8 \columnwidth]{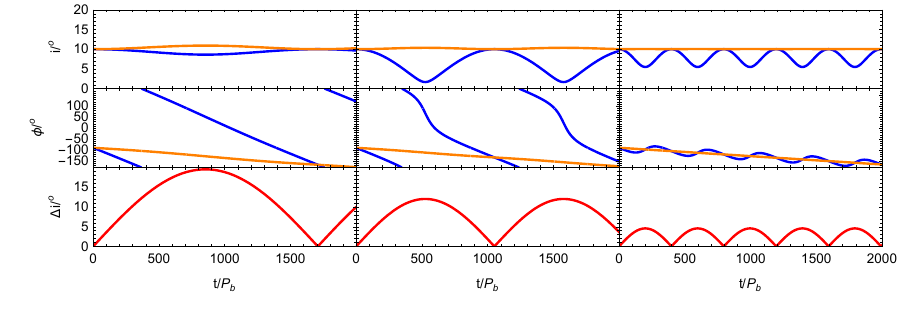}

\textbf{SPH simulations}\par\medskip

\includegraphics[width=1.8\columnwidth]{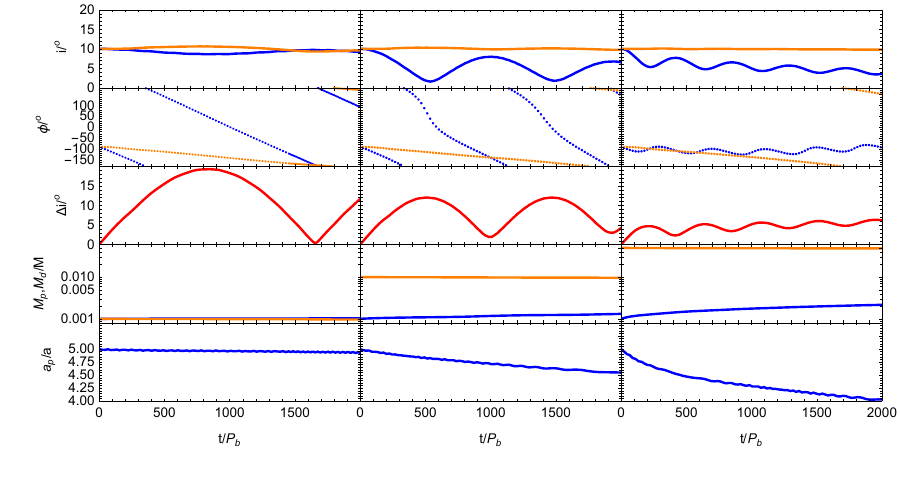}
 \includegraphics[width=0.33\linewidth]{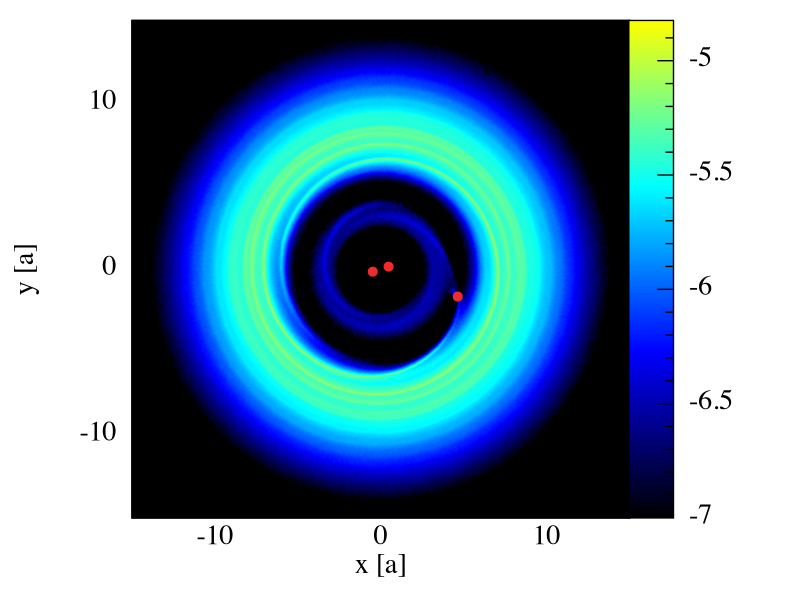}
     \includegraphics[width=0.33\linewidth]{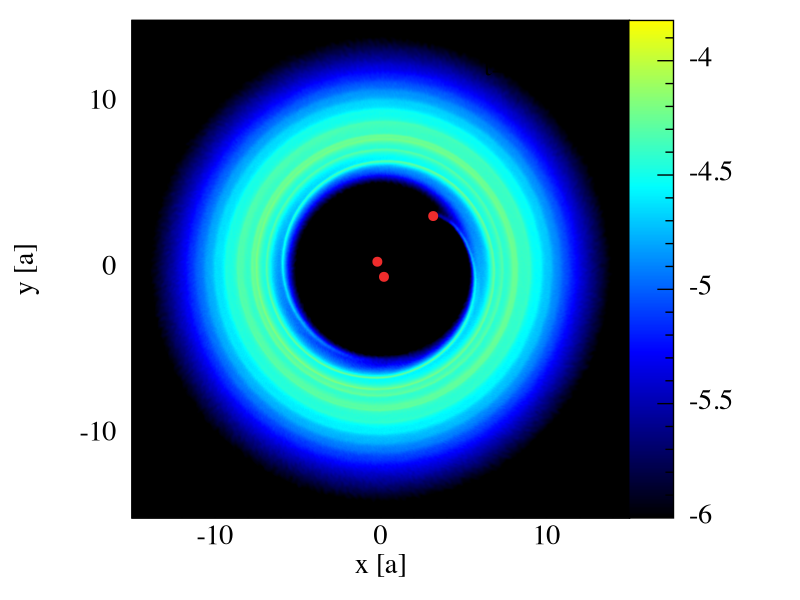}
     \includegraphics[width=0.33\linewidth]{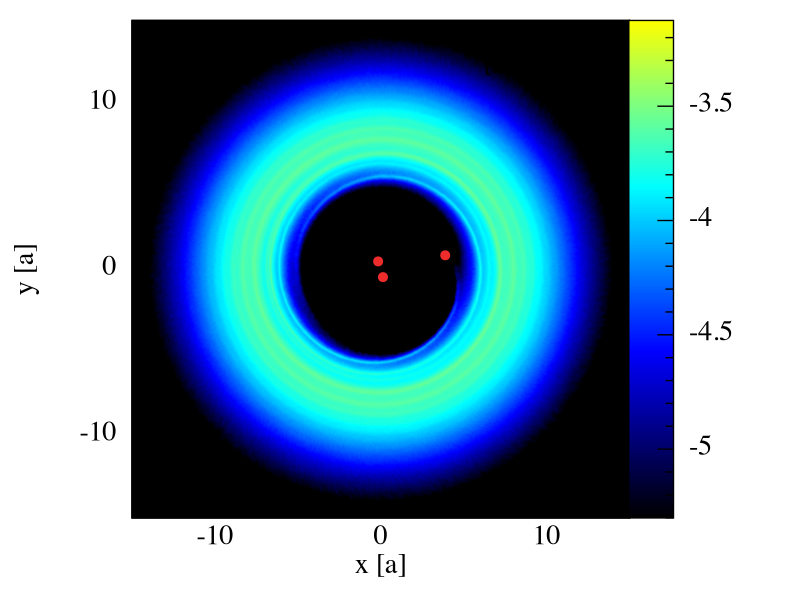}
\end{centering}
\caption{Evolution of a circumbinary disc  (orange lines) and planet of mass $0.001\,M$ (blue lines) around a
  circular orbit binary that is misaligned by $10^\circ$. The initial disc  mass is $M_{\rm d}=0.001\,M$ (left),  $0.01\, M$ (middle) and $0.05\,M$ (right). 
  The upper panel shows results for the analytic model in 
Section \ref{analytic} with $R_{\rm in}=6.5\,a_{\rm b}$ and $R_{\rm out}=12\,a_{\rm b}$ for the inclinations relative to the binary, the phase angles, and the relative
  inclination of the disc and the planet.
  The lower panels
  show the results of SPH simulations for the inclinations relative to the binary, the phase angles, the relative
  inclination of the disc and the planet, the masses and the
  separation of the planet from the centre of mass of the binary, respectively. The initial disc  mass is $M_{\rm d}=0.001\,M$ (left, circ1),  $0.01\, M$ (middle, circ2) and $0.05\,M$ (right, circ3).   The disc and planet masses overlap in the left panel and only the disc mass is visible.  The lower panel shows the column density of the disc in the initial frame of the binary orbit (the $x-y$ plane), at the end of each simulation. The red points show the binary stars and the planet with their radii three times larger than the sink size. The units of the column density are $Ma^{-2}$ with yellow being about two orders of magnitude larger than blue in each panel, but the absolute values are different between the three panels.  
  }
\label{sph1}
\end{figure*}

We first consider a low mass disc with a planet of mass $M_{\rm p}=M_{\rm d}=0.001\, M$. The
planet and the disc are initially coplanar to each other, but
misaligned to the binary orbit by $i_0=10^\circ$. Fig.~\ref{sph1} shows the analytic model in the upper panel and the SPH simulations in the two lower panels.   The left column of the middle panel of Fig.~\ref{sph1}
shows the evolution in time of the planet and the disc (circ1). For the low
mass disc, the planet and the disc are circulating. They precess on
independent timescales. However, there are tilt oscillations between
the two, as described in Section~\ref{analytic}.  The middle column of the middle panel of Fig.~\ref{sph1} shows a higher disc mass of $M_{\rm d}=0.01\,M$ (circ2), and the behaviour is qualitatively the same. However, the tilt oscillations of the planet have a much larger amplitude. Thus, the higher disc angular momentum causes the planet to be on average closer to coplanarity to the binary. 

The right column of the middle panel of Fig.~\ref{sph1} shows the evolution with a higher mass disc, $M_{\rm d}=0.05\,M$ (circ3). The planet and the disc initially precess on the
same average timescale, thus they are librating with respect to each
other.  
The SPH simulations in the middle panel of Fig.~\ref{sph1} show that, as predicted by the analytic results in
Section~\ref{analytic}, there is a critical mass above which the disc and the
planet librate and below which they circulate.

In the column density plots in the lower panel of Fig.~\ref{sph1} we see that a low density inner ring has formed in the low mass disc case (left). When the planet and the disc are circulating relative to each other, they spend less time in contact compared to when they librate relative to each other. Therefore material is more easily able to flow past the planet for lower disc masses.  

We apply the analytic theory in Section \ref{analytic} to models
with the same parameters as in Fig.~\ref{sph1}. The disc extends from an inner radius of $6.5\,a$ up to $12\,a$.
The results shown in the upper panel of Fig.~\ref{sph1} plot the corresponding results for the top three rows of the three SPH models plotted 
in the middle panel in Fig.~\ref{sph1}. The analytic model assumes that the planet mass and semi-major axis are constant in time and therefore we do not plot them for the analytic model.

The analytic results for the 
case of $M_{\rm d}=0.001 M$ (left column) are in quite good agreement with the SPH results. The models agree that the planet-disc system
undergoes nodal phase circulation.
For the case disc mass $M_{\rm d}=0.01 M$
(middle column), the analytic results  agree fairly well with the SPH results.  In the SPH results the inclinations are not quite
periodic and decay in time, while periodic behavior is predicted by the analytic model for a circular orbit binary. Again, the models agree that the planet-disc system
undergoes nodal phase circulation. The nonperiodic behaviour is a consequence of the disc
density evolution and the planet inward migration that is not accounted for in the analytic model.
For the high disc mass case with $M_{\rm d}=0.05 M$
(right column), there is rough agreement between the analytic and the SPH results. 
 The more nonperiodic behaviour is a consequence of the disc
density evolution of the more massive disc and the larger  
inward planet migration.
 But the periods of the planet inclination oscillations are in good agreement.
In addition, the models agree that the planet-disc system
undergoes nodal phase libration.

\subsection{High inclination disc}

\begin{figure*}
\begin{centering}
\includegraphics[width=1.8\columnwidth]{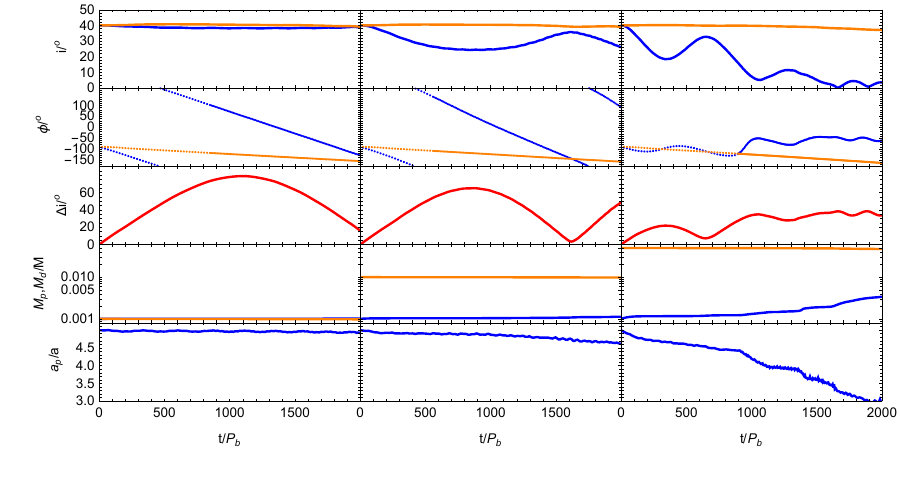}
  \includegraphics[width=0.33\linewidth]{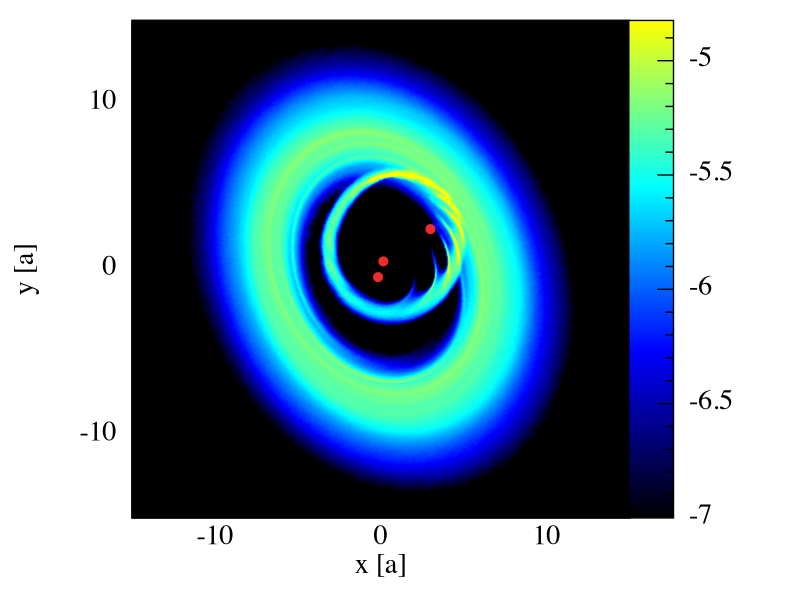}
        \includegraphics[width=0.33\linewidth]{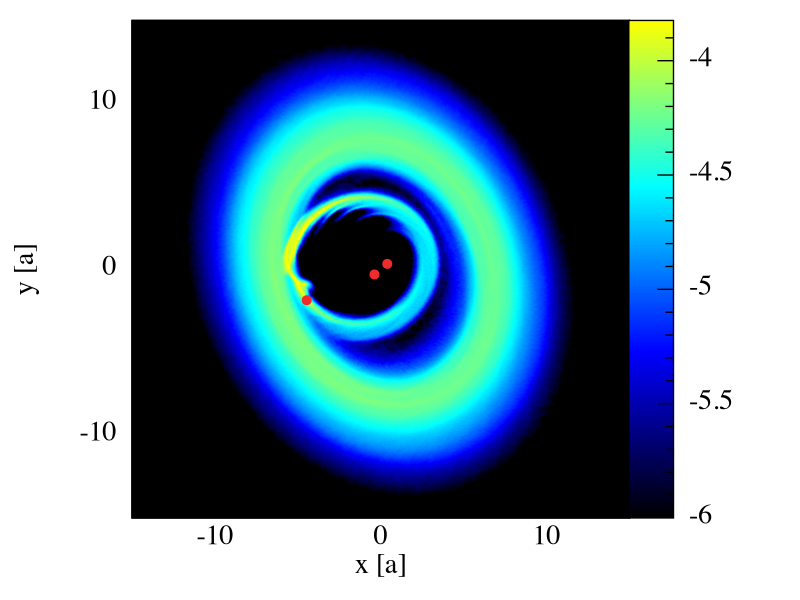}
          \includegraphics[width=0.33\linewidth]{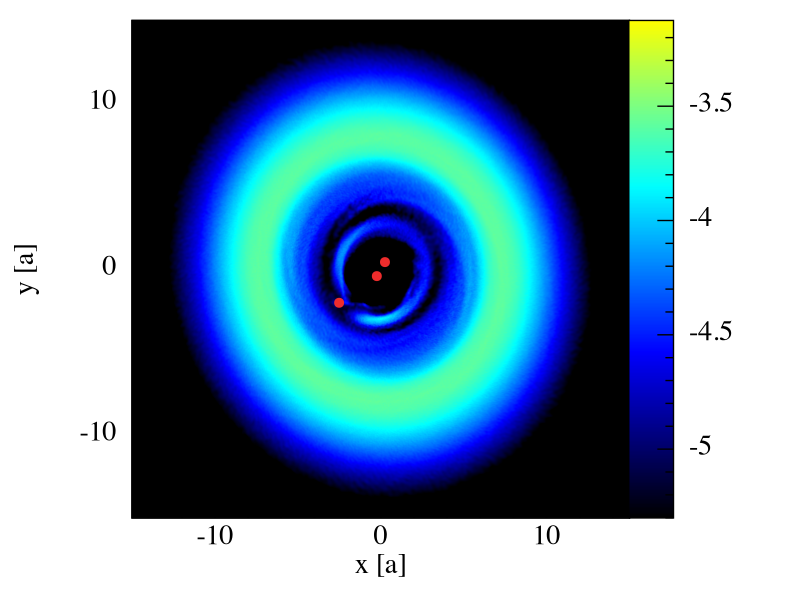}
\end{centering}
\caption{Same as  the upper and middle panels of Fig.~\ref{sph1} except the initial inclination of the planet and the disc is $40^\circ$. The initial disc masses are $M_{\rm d}=0.001\, M$ (left circ4),  $0.01\,M$ (middle, circ5) and $0.05\, M$ (right, circ6). The lower panel shows column density plots in the initial orbital plane of the binary (the $x-y$ plane).  }
\label{sph3}
\end{figure*}

Fig.~\ref{sph3} shows the evolution of the same systems as the SPH simulations in  Fig.~\ref{sph1} but with a higher initial inclination of $40^\circ$ (circ4, circ5 and circ6). The behaviour is
qualitatively the same in each case compared to the low inclination cases. 
While the
analytic results in Section~\ref{analytic} are valid only for small initial tilt, the results are qualitatively the same for higher initial inclination.

For the two lower disc masses,  the column density plots in Fig.~\ref{sph3} show that there is a relatively massive inner ring that has formed. The narrow ring has aligned close to coplanar to the binary orbit and has become highly eccentric, likely through its interaction with the circular orbit binary \citep[e.g.,][]{Shi2012, Munoz2020}. Material is able to flow past the planet more easily for the planet-disc system that is misaligned by $40^\circ$ compared to the system misaligned by $10^\circ$ because the mutual misalignment of the planet and the disc  oscillates to higher values.

\section{Eccentric orbit binaries}
\label{eccentric}

In this section, we use the same method as the previous section but we
increase the initial eccentricity of the binary up to $e_{\rm b}=0.5$.  We
consider a total of 11 simulations covering three different disc masses and various different initial inclinations  as shown in
Table~\ref{tab} (ecc1 - ecc11). We first discuss the orbital evolution of an isolated circumbinary planet and an isolated circumbinary disc
around an eccentric binary so that the interaction between the disc and the planet  is clear.

\subsection{Evolution of an isolated circumbinary planet}
\label{planet}  

For low inclination, the planet angular
momentum precesses about the binary angular momentum vector (as is the
case for a circular orbit binary). However, for higher inclination, the
planet precesses about the direction of polar alignment. For a low mass planet, the direction of polar
alignment is along to the eccentricity vector of the binary (binary semimajor axis).  For
higher mass planet, the generalised polar alignment direction is at a
lower level of misalignment with respect to the binary \citep{MartinLubow2019,Chen2019}.  

The minimum critical initial inclination angle above which the planet
librates with respect to the binary depends only the binary
eccentricity and the angular momentum ratio of the planet to the
binary, $j$. 
The critical minimum critical inclination is 
\begin{equation}
    \cos{i_{\rm min}} = \frac{\sqrt{5} e_{\rm b} \sqrt{4 e_{\rm b}^2 -4j^2(1-e_{\rm b}^2)+1}- 2j(1-e_{\rm b}^2)}{1+4e_{\rm b}^2}.
    \label{minlib}
\end{equation}
for $\chi>0$, where
\begin{equation}
\chi=e_{\rm b}^2 - 2 (1-e_{\rm b}^2) j (2 j + \cos{i}).
\label{lambda}
\end{equation}
This corresponds to small $j$ or  large $e_{\rm b}$. On the other hand,  the minimum critical inclination is given by
\begin{equation}
    \cos{i_{\rm min}} = \frac{ \sqrt{(1- e_{\rm b}^2) \left(1+4e_{\rm b}^2 +60 (1-e_{\rm b}^2) j^2 \right)} -(1 - e_{\rm b}^2) }{10 (1-e_{\rm b}^2) j}
    \label{minlib2}
\end{equation}
for $\chi<0$ and this corresponds to large $j$ or small $e_{\rm b}$ \citep{MartinLubow2019}.
For a planet of mass $0.001\,M$ at orbital separation of
$5\,a$, the angular momentum ratio of the planet to the binary is
$j=0.01$. The critical inclination is $38.5^\circ$ for binary
eccentricity $e_{\rm b}=0.5$ (using equation~(\ref{minlib})).  Above this inclination, a planet librates,
while below it circulates. Thus, simulations in which the planet begins with
$i>38.5^\circ$ (ecc4-ecc9) would librate in isolation (meaning without the presence of the disc). However, as shown by the 7th column in Table~\ref{tab}, this is affected through interactions with the disc that we examine here.

\subsection{Evolution of an isolated circumbinary disc}
\label{disc}

A higher disc mass leads to a larger critical
inclination required for libration of the disc in the absence of the planet. The initial angular
momentum of the disc compared to the binary is about $j=0.026$, $j=0.26$ and $j=1.3$ for disc masses $M_{\rm d}=0.001\, M$, $M_{\rm d}=0.01\, M$ and $M_{\rm d}=0.05\, M$, respectively. The corresponding minimum critical inclination angles are $40^\circ$ (calculated with equation~(\ref{minlib})), $52^\circ$ and $45^\circ$ (both calculated with equation~(\ref{minlib2})). 
Thus, in isolation, without the presence of the planet, we would expect the disc and binary to be circulating in runs ecc1-ecc6 and librating for runs ecc7-ecc9 in the 8th column in table~\ref{tab}.

\subsection{Low inclination planet-disc system}

\begin{figure*}
\begin{centering}
\textbf{Analytic model}\par\medskip

\includegraphics[width=1.8\columnwidth]{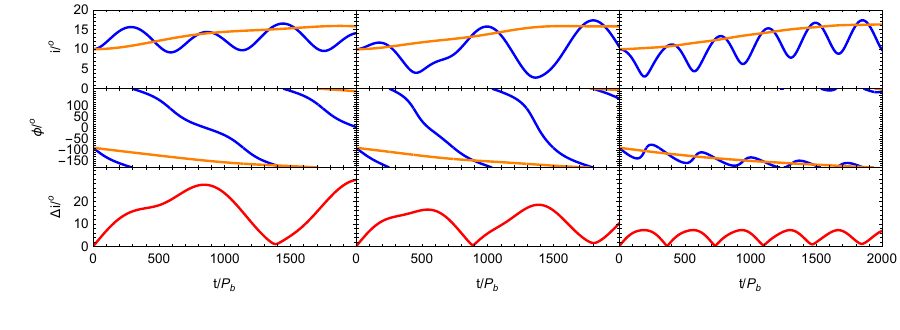}

\textbf{SPH simulations}\par\medskip

\includegraphics[width=1.8\columnwidth]{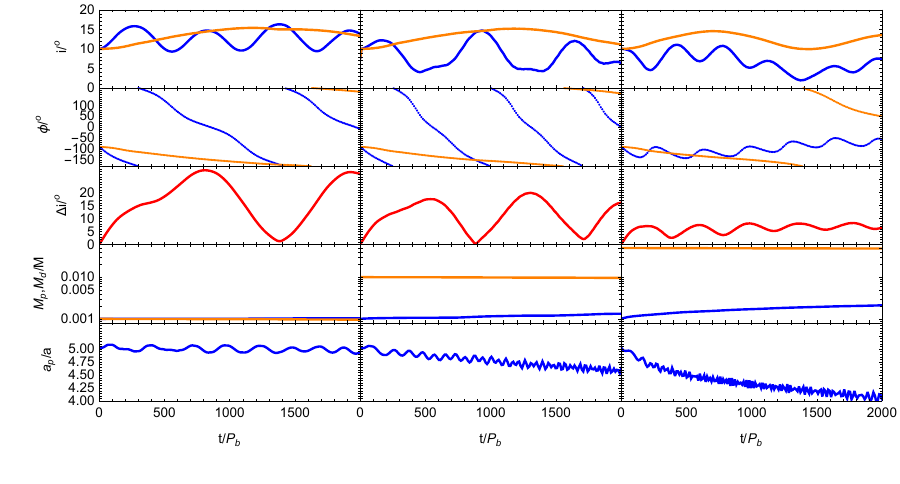}
   \includegraphics[width=0.33\linewidth]{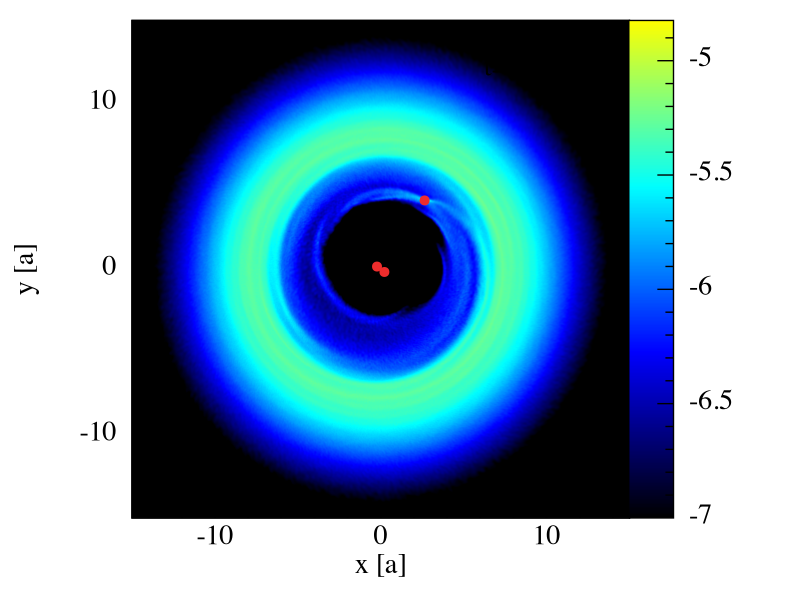}
        \includegraphics[width=0.33\linewidth]{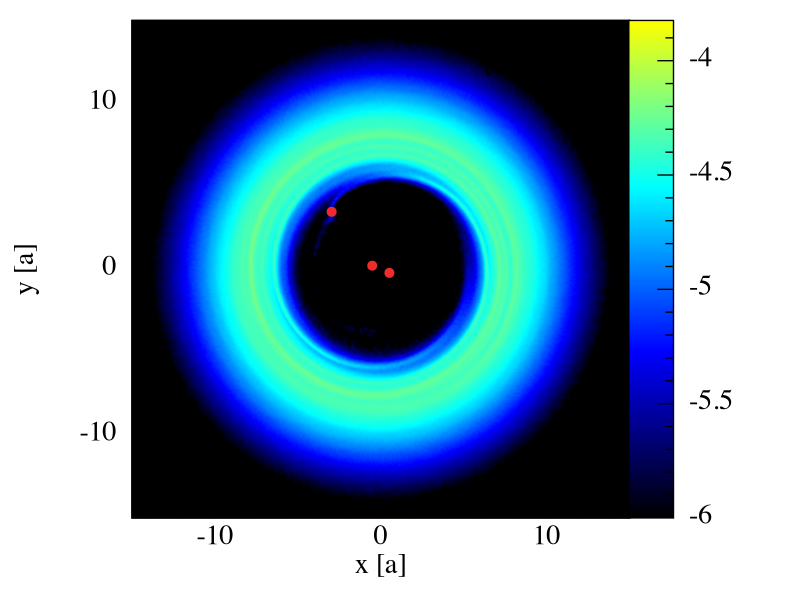}
          \includegraphics[width=0.33\linewidth]{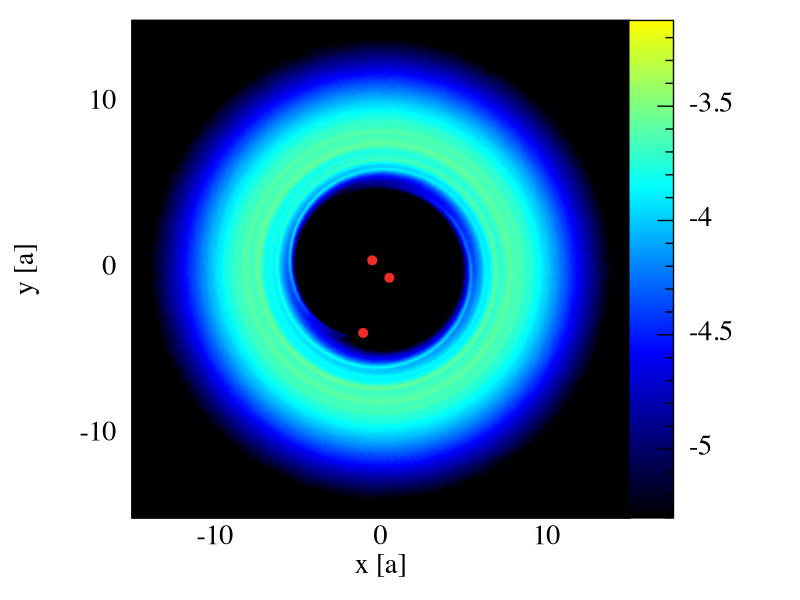}

\end{centering}
\caption{Same as Fig.~\ref{sph1} except the initial binary
  eccentricity is $e_{\rm b}=0.5$.  The upper panel is  based on the analytic model in 
Section \ref{analytic} with $R_{\rm in}=6.5\,a_{\rm b}$ and $R_{\rm out}=12\,a_{\rm b}$. The lower panels show the results of SPH simulations. The initial disc masses are
  $0.001\,M$  (left, ecc1),  $0.01\,M$ (middle, ecc2) and $0.05\, M$ (right, ecc3).
  }
\label{sph2}
\end{figure*}

Fig.~\ref{sph2} shows the evolution of discs around an eccentric orbit binary with $e_{\rm b}=0.5$ for a low initial inclination of $10^\circ$. The upper panel shows the analytic model and the lower two panels show the SPH simulations. Both the disc and the planet are well within the circulating regime, meaning that they would precess about the binary angular momentum vector, as they would if they did not interact with each other. For an eccentric orbit binary, there are still inclination oscillations even if the planet and disc do not interact because of the non-axisymetric potential of the eccentric orbit binary \citep{Smallwood2019}. In the simulation with lowest disc mass of $0.001\, M$ (left column of the middle panel), the disc and the planet  undergo tilt oscillations driven by the binary, almost independently of each other. However, in the higher-mass cases (middle and right columns of the middle panel), the planet inclination does not increase initially and instead drops.
There are now competing effects that drive tilts oscillations:
effects of the binary on the planet and disc and the effects of the planet and
disc on each other. The latter effect becomes more important at higher disc masses, whereas the former effect is independent of disc mass.
For the lowest mass disc case in Fig.~\ref{sph2}, the planet inclination initially rises, unlike the case of a circular orbit
binary in Fig.~\ref{sph1}, suggesting that the eccentric orbit binary
is dominating the planet inclination evolution.
While for the highest mass disc case in Fig.~\ref{sph2}, the planet inclination initially declines, like the case of a circular orbit
binary in Fig.~\ref{sph1}, suggesting that the disc
is dominating the planet inclination evolution.
For low-mass discs, the planet and the disc undergo tilt oscillations independently of each other leading to high-inclination planets. For high-mass discs the secular tilt oscillations dominate and the planet inclination is on average closer to alignment with the binary, as
the disc evolves to alignment with the binary.

We apply the analytic theory in Section \ref{analytic} to models
with the same parameters as in the lower panels of Fig.~\ref{sph2}.
The results shown in the upper panel in Fig.~\ref{sph2} plot the corresponding results for the inclination, phase angle, and mutual planet-disc inclination.
For an eccentric orbit binary, the tilt oscillations are generally non-periodic, even if the effects of disc evolution and planet migration are ignored as discussed in Section~\ref{analytic}.

The analytic results for the
case of $M_{\rm d}=0.001 M$ (left column) are generally in agreement.
The planet inclination initially increases in both models.  This initial increase is a consequence of the effects of the eccentric binary potential for the chosen initial nodal phase $\phi= -90^\circ$. For an initial value of $\phi=0^\circ$, the planet inclination would initially decrease. For the first three planet tilt oscillations, the second maximum is the smallest, with the first maximum the next largest, and the third maximum being the largest. This pattern is found in both the analytic and the SPH results.  
For the case of disc mass $M_{\rm d}=0.01 M$
(middle column), the reduced initial increase in the inclination of the planet relative to the case $M_{\rm d}=0.001 M$ is found
in both the analytic and SPH results. The subsequent planet inclination evolution is quite complex and there is rough agreement between the analytic and SPH results.
The complicated nonperiodic behaviour is a consequence of binary eccentricity that is accounted for in the analytic model, as well as the disc
density evolution and the planet inward migration that is not accounted for in the analytic model.
For the disc mass $M_{\rm d}=0.05 M$ case
(right column), the planet inclination initially declines
in both the analytic and SPH results. Both models
roughly agree on the initial period of the relative inclination
of about $400 P_{\rm b}$. 
The SPH results suggest that the relative inclination does not
go to zero in time. 
The analytic and SPH models agree that the planet-disc system
initially undergoes mutual nodal libration. 

\begin{figure*}
\begin{centering}
\includegraphics[width=1.8\columnwidth]{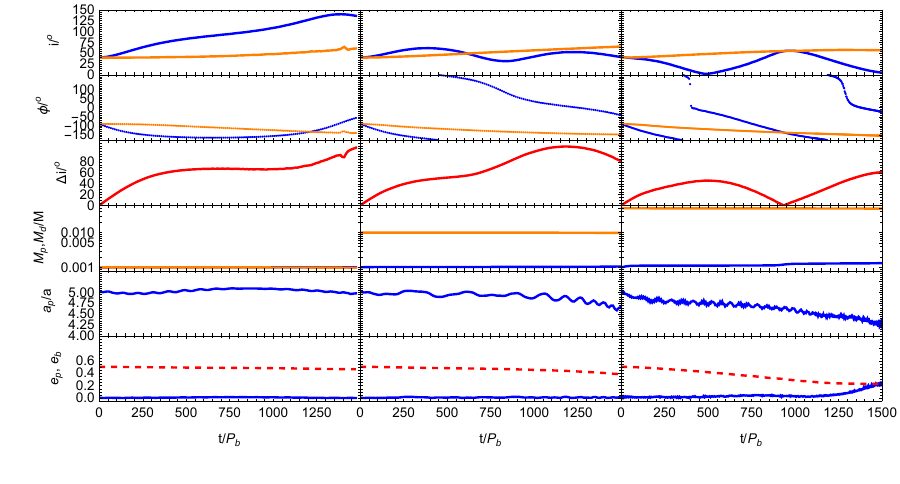}
\includegraphics[width=0.33\textwidth]{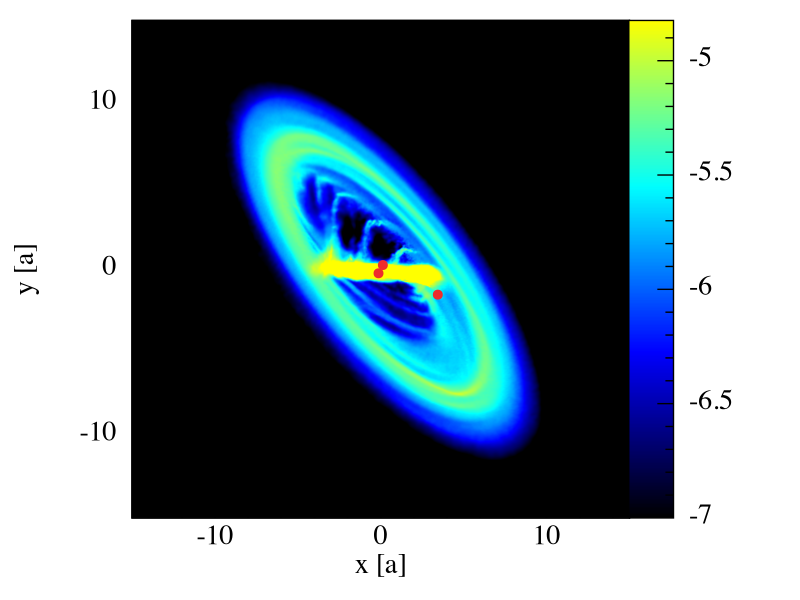}
\includegraphics[width=0.33\textwidth]{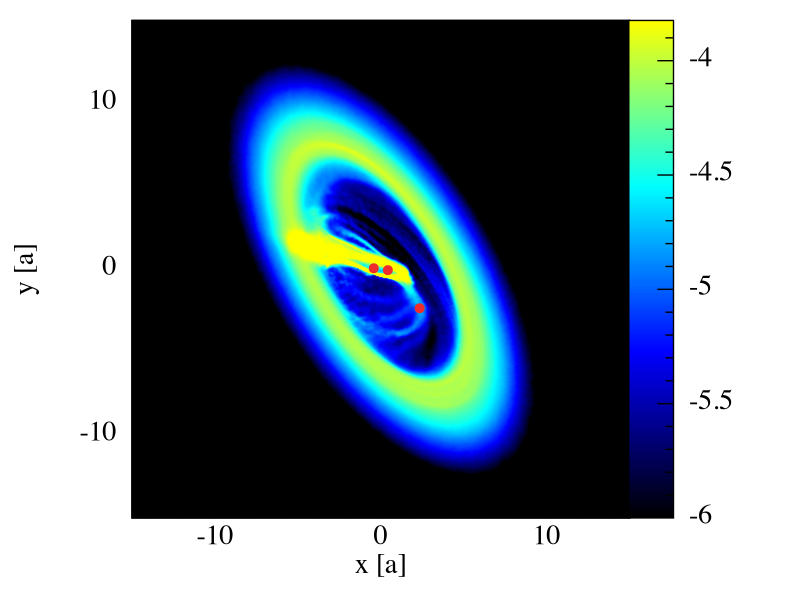}
\includegraphics[width=0.33\textwidth]{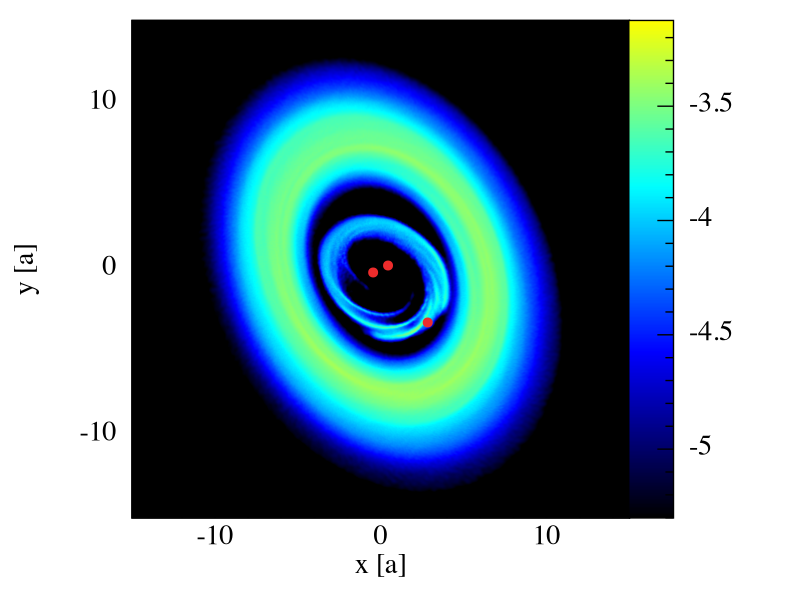}
\end{centering}
\caption{Same as Fig.~\ref{sph1} except the initial inclination of the planet and the disc is $40^\circ$ and the  initial binary
  eccentricity is $e_{\rm b}=0.5$ and there is an additional lower row to show the eccentricity of the planet (blue line) and the binary (red dashed line). The initial disc masses are
  $0.001\,M$  (left, ecc4),  $0.01\,M$ (middle, ecc5) and $0.05\, M$ (right, ecc6). The column density plots are shown in the $x-y$ plane. The  lower left and middle plots have inner polar disc rings while the lower right plot has an inner coplanar ring.
  }
\label{sph7}
\end{figure*}

\subsection{Planet-disc system with initial inclination $i=40^\circ$}

Fig.~\ref{sph7} shows a planet-disc system with
initial inclination $40^\circ$ and initial binary eccentricity $e_{\rm b}=0.5$. In isolation, the disc in all cases is close to or below the critical inclination for libration relative to the binary (see Section~\ref{disc}). In all disc mass cases, the disc is circulating with respect to the binary. The disc inclination is increasing, at least initially, in all cases because of the tilt oscillations associated with the precession in the eccentric binary.  

Since the planet has a lower angular momentum than the disc, in isolation (without the disc) it is above the critical angle required for libration.  For the low disc mass $M_{\rm d}=0.001\,M$ (left panel, ecc4) the planet does librate as the planet does not interact strongly with the disc. However, for the higher mass disc cases  $M_{\rm d}=0.01\,M$ (middle, ecc5) and $M_{\rm d}=0.05\,M$ (right panel, ecc6)
there is a stronger interaction between the planet and the disc and the planet is circulating with respect to the binary, rather than librating.  This is because the planet and the disc interact strongly
enough for  secular tilt oscillations to be important.  Thus, a high mass disc that is circulating is able to prevent a planet from librating with respect to the binary and the average inclination of the planet becomes closer to coplanar. 

The column density plots in Fig.~\ref{sph7} show that in each case there is a dense inner ring that has formed. This is because there is a large mutual misalignment between the planet and the disc and material is able to flow past the planet. The inner ring for the lower two disc masses has aligned to a polar orientation to the binary, while it is coplanar in the highest disc mass case. At earlier times in the high mass case, the ring is also polar. Complex evolution of the inner ring can be driven by the competition between the inner binary and the outer massive disc \citep{Martin2022}. The binary drives circulation or libration relative to the binary that leads to coplanar or polar alignment, respectively. The massive outer disc  can drive Kozai-Lidov \citep[KL,][]{vonZeipel1910,Kozai1962,Lidov1962} oscillations of the inner disc and it moves towards coplanar alignment \citep{Martinetal2014}. The process of KL oscillations driven
by an outer disc around a single star was first described by \cite{Terquem2010}. 
KL oscillations around a single star 
 involve a nearly constant planet semimajor axis $a_{\rm p}$. The 
reason is that central star and disc provide a static potential so that planet orbital energy that depends on $a_{\rm p}$ is  conserved.
However, in this case, the KL oscillations do not involve a static potential
mainly because of the central binary.
These simulations run very slowly because of the dense inner ring and therefore have a shorter end time of $t=1500\,P_{\rm b}$ compared to the other simulations in this work.  

A massive circumbinary disc can also drive KL oscillations of the planet-binary orbit and/or the binary orbit if their orbits are sufficiently  misaligned to the disc \citep[e.g.][]{Terquem2010,Huang2025}.  In the high mass disc case in Fig.~\ref{sph7},  the planet eccentricity starts to increase at the end of the simulation. If the simulation was run for longer this may play a larger role in the evolution. However, KL oscillations in this case may be limited to relatively low magnitude because the mutual inclination between the planet and the disc is on average low. KL oscillations are also occuring in the binary and the binary eccentricity has decreased significantly by the end of the simulation.  

\subsection{Planet disc system with initial inclination $i=60^\circ$}

\begin{figure*}
\begin{centering}
\includegraphics[width=1.8\columnwidth]{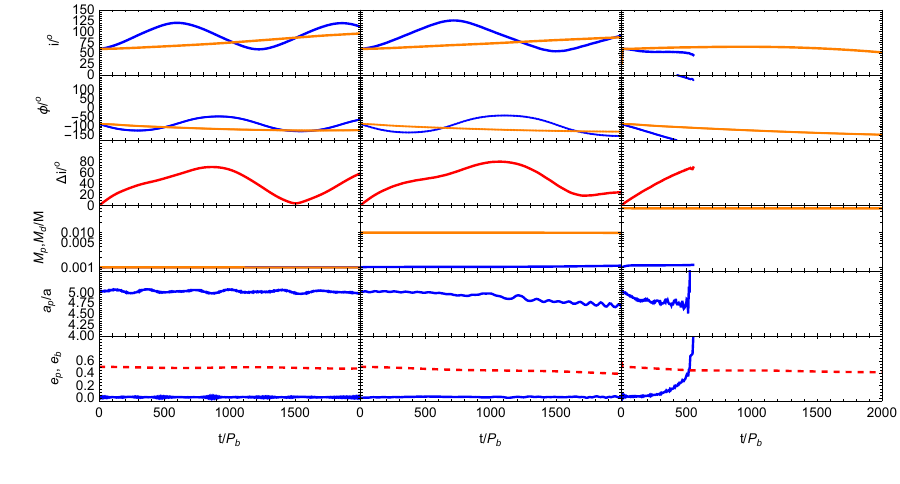}
\includegraphics[width=0.66\columnwidth]{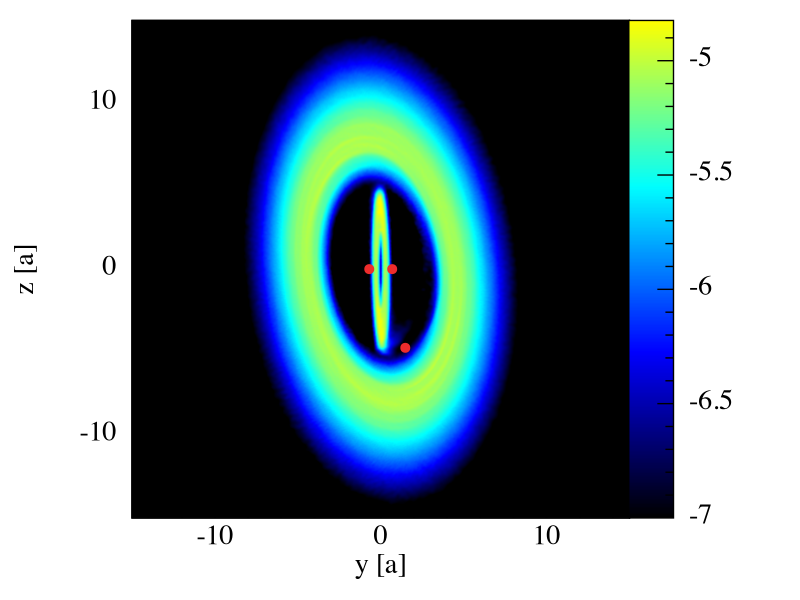}
\includegraphics[width=0.66\columnwidth]{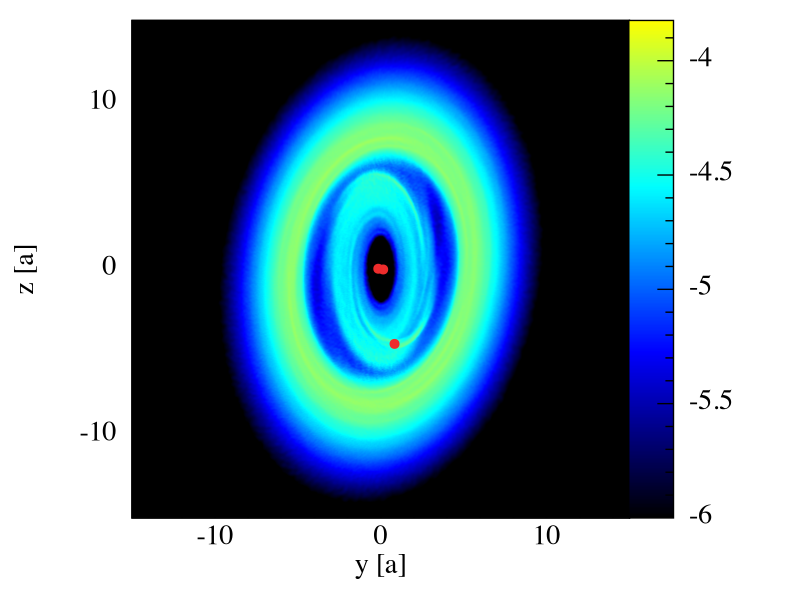}
\includegraphics[width=0.66\columnwidth]{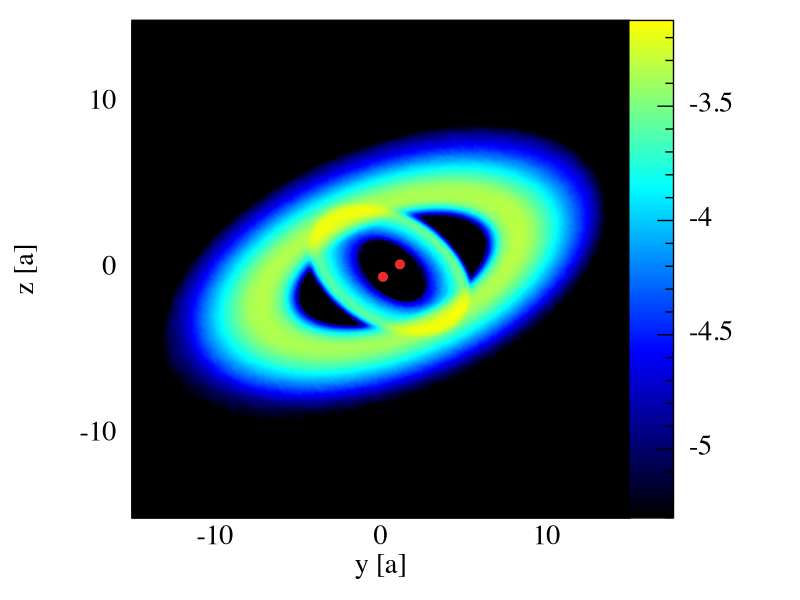}
\end{centering}
\caption{Same as Fig.~\ref{sph1} except the
  initial inclination of the planet and the disc is $60^\circ$ and the initial binary eccentricity is $e_{\rm
    b}=0.5$ and there is an additional lower row to show the eccentricity of the planet (blue line) and the binary (red dashed line).}  The initial disc masses are
  $0.001\,M$ (left, ecc7),  $0.01\,M$ (middle, ecc8) and $0.05\, M$ (right, ecc9). The column density plots are shown in a frame perpendicular to the initial binary orbit (the $y-z$ plane). 
\label{sph9}
\end{figure*}

 Fig.~\ref{sph9} shows a system that begins with a higher inclination $i=60^\circ$ around an eccentric binary. In isolation, both components would undergo libration.  
For the disc masses $M_{\rm
  d}=0.001\,M$ (left, ecc7) and $M_{\rm d}=0.01\,M$ (middle, ecc8), both the disc and the planet
librate with respect to the binary and there is little interaction
between the planet and the disc. They librate on different
timescales because of their different orbital radii. Planets that form in highly misaligned discs may be left in a highly noncoplanar orbit with respect to the binary orbit, after the disc has dissipated.

The right panel of Fig.~\ref{sph9} shows a disc with high mass $M_{\rm d}=0.05\,M$
(ecc9). In this case we would expect the planet to librate with the disc because the gravitational coupling between them is even stronger than the case of the $0.01 M$ mass disc in the center panel of Fig.~\ref{sph9}. However, as we describe below, that is not the case, due to a major change in the planet's orbit during its evolution.
Initially, the planet and the disc
precess  at somewhat different rates and once they become sufficiently mutually misaligned,
the planet undergoes a KL oscillation. Consequently, the
orbit becomes highly eccentric.  
The planet
has a close encounter with one component of the binary at a time of
about $500\,\rm P_{\rm b}$ and gets ejected from the system. This has been seen before in simulations of misaligned multiplanet systems around binaries \citep{Chen2023stability,Chen2023three,Chen2024ffps}.

In these highly tilted simulations the column density plots show inner polar rings for the two lower mass discs. Interestingly, the higher mass disc case still has a broken disc, even though the planet has been ejected. After the ejection of the planet, the disc rings undergo nodal precession relative to the binary orbit on different timescales. Over time, the rings may viscously spread and join together again \citep[e.g.][]{Smallwood2021}.

\subsection{Planet disc system with initially near polar inclination}

\begin{figure*} 
\begin{centering}
\includegraphics[width=1.8\columnwidth]{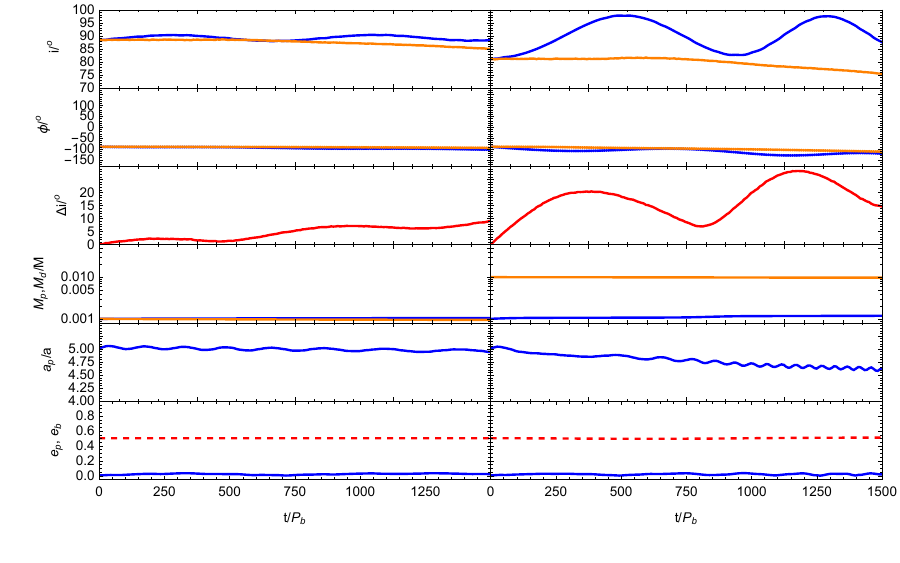}
\includegraphics[width=0.8\columnwidth]{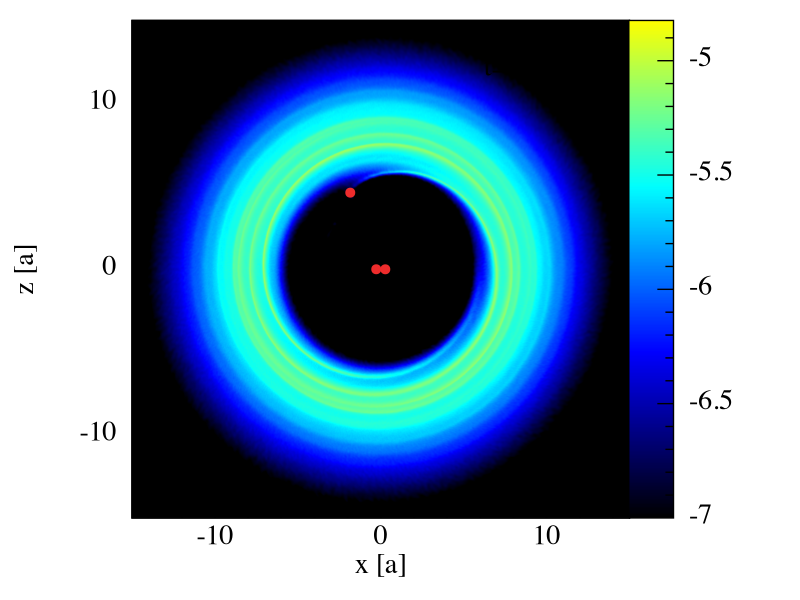}
\includegraphics[width=0.8\columnwidth]{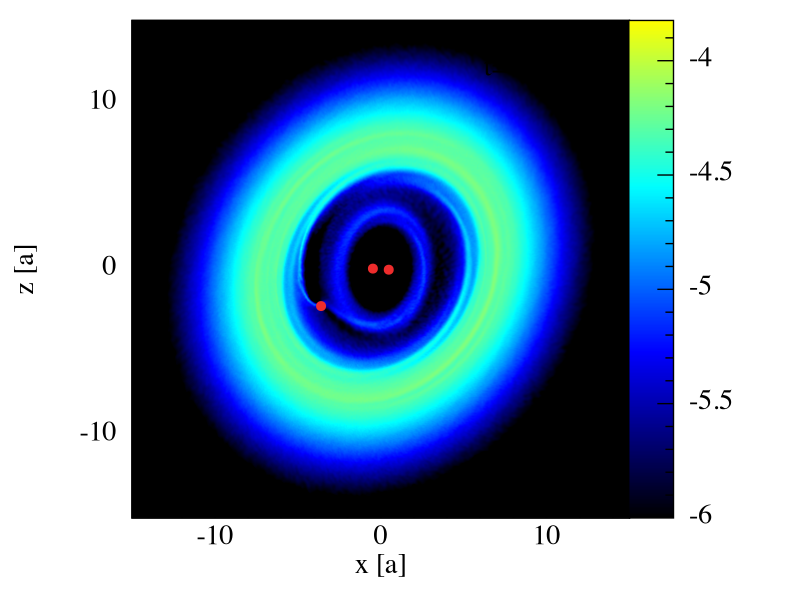}
\end{centering}
\caption{Same as Fig.~\ref{sph1} except the
  initial inclination of the planet is the generalised polar inclination and  the initial binary eccentricity is $e_{\rm
    b}=0.5$ and there is an additional lower row to show the eccentricity of the planet (blue line) and the binary (red dashed line).  In the left panel, the initial inclination is $88.5^\circ$ with disc mass $M_{\rm d}=0.001\, M$ (ecc10) and in the right panel the initial inclination is $81.2^\circ$ and the disc mass is $M_{\rm d}=0.01\,M$ (ecc11). The column density plots are in the $x-z$ plane.}
\label{sph10}
\end{figure*}

The inclination of the stable polar state around an eccentric binary depends upon the eccentricity of the binary and the angular momentum of the circumbinary objects. The polar state is $90^\circ$ for a test particle, but at lower levels of misalignment for a third body with mass \citep{MartinLubow2019, Chen2019}. We consider here a planet-disc system that begins in the generalised polar state. 
The stationary inclination is given by
\begin{equation}
\cos i_{\rm s} =  \frac{ -(1+ 4e_{\rm b}^2)+ \sqrt{ \left(1+ 4e_{\rm b}^2 \right)^2+60 (1-e_{\rm b}^2) j^2} }{10 j}
\end{equation}
\citep{MartinLubow2019}.

For $j$, we use the sum of the angular momentum of the planet and the disc. For our three disc masses $M_{\rm d}=0.001\,M$, $M_{\rm d}=0.01\, M$ and $M_{\rm d}=0.05\, M$ we find the misalignment to be $88.5^\circ$, $81.2$ and $64.0^\circ$, respectively. Fig.~\ref{sph10} shows the two lower disc mass simulations that begin at the generalised polar state.\footnote{\cite{Chen2022} carried out a more precise determination of the polar stationary inclination
that included the effects of the gravitational interaction between both orbiting bodies which were both particles. Based on those results, we expect that the stationary configuration actually involves  somewhat different inclinations for  the planet and disc. However, this is beyond the scope the current work.} We do not consider the massive disc case here  since the stationary inclination is close to the $60^\circ$ case shown in ecc9. 
Fig.~\ref{sph10} shows the results for simulations that begin at these inclinations. In both cases, the planet and disc undergo  growing mutual inclinations. The planet is on average closer to a $90^\circ$ misalignment to the binary orbit than the disc. The planet-disc interactions therefore can increase the inclination of the planet relative to the binary. 
The column density plots show that the planet region is mostly clear, but a low density inner ring forms in the higher mass disc case as a result of the larger planet-disc mutual tilt.   
The planet and binary eccentricities do not show significant changes. 
 
\section{Conclusions}
\label{concs}
 
We have explored the evolution of planet--disc systems around a
misaligned and eccentric binary star with analytic secular theory and
hydrodynamic simulations. The planet, disc and binary interact through
gravitational forces. The planet has a mass high enough to open a gap
in the circumbinary disc.  The disc is taken to lie initially somewhat outside the orbit of the planet. The planet and the disc are initially
coplanar to each other but misaligned to the binary orbit. The planet
and the disc do not generally remain coplanar to each other. The initial tilt relative to the binary orbit represents the tilt state at which the giant planet forms. Initially nearly coplanar or polar models represent cases in which the disc forms either polar or coplanar. Alternatively it can represent cases in which planet formation occurs after significant tilt evolution towards polar or coplanar states.

For a circular orbit binary, the planet and the disc both always
precess about the binary angular momentum vector. The circular orbit of the binary results in a binary secular potential that is axisymmetric about its angular momentum vector.  As a result, the binary does not directly cause the planet and disc to undergo changes in tilt relative to the binary orbital plane. Changes in tilt occur due the planet-disc interactions. 
A giant planet and a circumbinary disc do not remain
coplanar but instead undergo secular mutual tilt oscillations.  The average
inclination of the planet relative to the binary is lower than the average inclination of the
disc, as was also found by \cite{Pierens2018}. Above a
critical disc mass, the planet and the disc precess at the
same average rate (i.e., undergo mutual libration) with a more limited mutual 
inclination oscillation  amplitude.

For eccentric orbit binaries, if the system is initially sufficiently
misaligned it may precess about a generalised polar state rather than
the coplanar state. The eccentricity of the binary causes the binary secular potential to be nonaxisymmetric. Due to the binary potential, the planet and disc undergo secular tilt oscillations relative to the binary as they nodally precess.  
Additional changes in tilt occur due to planet-disc interactions. 
 The planet and disc initial nodal phase angles
are taken to be $\phi = -90^{\circ}$ relative to the binary eccentricity vector.
For a low disc mass at high inclination, the planet
and the disc librate and undergo tilt oscillations on different timescales with little
interaction between the two. 
With the initial phase angle $\phi = -90^{\circ}$, the planet and disc initially undergo an 
increase in tilt relative to the binary orbital
plane.
However, for a high mass disc, the system
may initially undergo a secular tilt oscillation leading the
inclination of the planet to decrease initially. Thus, the range of initial inclinations for which the planet librates  relative to the binary is reduced through the planet-disc interaction.   Futhermore, a massive disc can lead to
Kozai--Lidov oscillations of the planet. A highly eccentric planet can
have strong interaction with the binary leading to it being flung out
to much larger radius, or possibly collision with one of the binary components.

For cases where the planet and the disc become highly mutually misaligned, material can easily flow past the planet and form a high density inner disc ring. The inner ring may undergo complex dynamics as a result of its interaction with the binary and the outer disc and planet. The ring can undergo  eccentricity growth driven by the binary \citep[e.g.][]{Shi2012,Thun2017,Munoz2020,Lubow2022} and eccentricity growth from KL oscillations driven by the outer massive disc \citep[e.g.][]{Terquem2010,Huang2025}. 
Observations of protoplanetary discs show large numbers of shadows \citep[e.g.][]{Marino2015,Benisty2017} that may be caused by inner disc rings  that are misligned to the outer disc \citep{Price2018b,Nealon2019}. Our simulations show that high density inner disc rings form for circumbinary discs with a giant planet,
but the orientation of the initial disc is not close to coplanar or to polar to the binary orbit. The binary-disc and binary-planet interactions can lead to long lived misaligned rings.  We do note that the disc temperature we have chosen is relatively cool with $H/R=0.02$. However, a larger disc aspect ratio may lead to the gap closing when the planet becomes highly misaligned \citep[e.g.][]{Smallwood2021}.

If the distribution of circumbinary planet inclinations is isotropic, it has been estimated that circumbinary planets are more abundant than planets around single stars \citep{Armstrong2014}. 
Observations and simulations suggest that circumbinary discs form with a broad distribution of initial inclinations \citep{Czekala2019, Elsender2023}. However, even if the circumbinary discs form with an isotropic misalignment,  the  final distribution of planet inclinations is not isotropic.  If the circumbinary disc is sufficiently warm and narrow enough to precesses as a single body, it may align to either coplanarity or polar alignment within the lifetime of the disc. Planets that form within the already aligned discs will have two inclinations, coplanar and polar. 
However, the inclination of planets that form in a misaligned disc can evolve. Around a circular binary the inclination of giant planets that form in the disc will be closer to coplanar alignment than the disc (see Figs.~\ref{sph1}-\ref{sph3}).  Thus, even if the discs are distributed isotropically, the planets around circular orbit binaries are more likely to be found close to coplanar. 
Around an eccentric binary, a massive disc causes the planet inclination to move towards either coplanar or polar, depending upon the initial disc inclination. 
However, for lower disc masses, planets may be found on highly misaligned orbits around eccentric binaries
(see Fig.~\ref{sph2}).

\section*{Acknowledgments} 

We acknowledge support from NASA through grants  80NSSC19K0443 and 80NSSC21K0395. Computer support was provided by UNLV's
National Supercomputing Center.
We acknowledge the use of SPLASH \citep{Price2007} for
the rendering of the figures.

\section*{DATA AVAILABILITY}
\label{da}
The data underlying this article will be shared on reasonable request to the corresponding author.

\bibliographystyle{mnras} 
\bibliography{martin}

\label{lastpage}
\end{document}